\documentclass[aps,superscriptaddress,nofootinbib,eqsecnum,prd,notitlepage,twocolumn]{revtex4-1} 

\pdfoutput=1

\usepackage{amsfonts}
\usepackage{amsmath}
\usepackage{amssymb}
\usepackage{graphicx,color}
\usepackage{float}
\usepackage{hyperref}
\usepackage{subfigure}
\usepackage{dcolumn}
\usepackage{soul}
\usepackage{ulem}
\usepackage{verbatim}

\begin{document}

\title{Symmetry breaking patterns for two coupled complex scalar fields
  at finite temperature and in an external magnetic field}

\author{R. L. S. Farias}
\affiliation{Departamento de F\'{\i}sica, Universidade Federal de
  Santa Maria, 97105-900
Santa Maria, RS, Brazil}

\author{Rudnei O. Ramos}
\affiliation{Departamento de F\'{\i}sica Te\'{o}rica, Universidade do
  Estado do Rio de
Janeiro, 20550-013 Rio de Janeiro, RJ, Brazil}

\author{D\'erick S. Rosa}
\affiliation{Instituto Tecnol\'{o}gico de Aeron\'{a}utica, DCTA,
  12228-900 S\~{a}o Jos\'{e} dos Campos, SP, Brazil}

\begin{abstract}

A model of two coupled complex scalar fields is studied at finite
temperature and under an external magnetic field. The results are
obtained in the context of the nonperturbative method of the optimized
perturbation theory and contrasted with those obtained in perturbation
theory and in the one-loop approximation.  The emergence of phenomena
related to inverse symmetry breaking and symmetry nonrestoration are
analyzed.

\end{abstract}

\maketitle

\section{Introduction}

Phase transitions are ubiquitous in nature. In physics, phase transition
phenomena range from those possibly
happening at high energies in the early Universe, e.g.,  at the
electroweak~\cite{Trodden:1998ym} and quantum chromodynamics (QCD)
scales~\cite{Schwarz:2003du,BraunMunzinger:2009zz}, to the ones in
lower-energy systems, like in condensed matter~\cite{Goldenfeld:1992qy}
and in atomic gas systems~\cite{Cornell:2002zz}.  The
first studies along this topic give a good qualitative and
quantitative idea of how these phenomena might occur.  The general
expectation is that a system initially in a symmetry broken state will
have its symmetry restored at sufficiently high  temperatures, beyond
some critical temperature. An unambiguous example is the rotational
symmetry of a ferromagnetic system, which is broken at temperatures
below a critical value, the Curie critical temperature, and then gets restored  
as the temperature is raised beyond that critical value.

The above picture gets less clear when more than one symmetry is at
play and multiple critical points can be present. This is a common
situation in condensed-matter systems, where reentrant (intermediate)
phases are present~\cite{reentrant}. In this case, the intermediate
phase might, e.g., be a state which is less symmetrical than the previous one and
later phases. This is a case we can characterize as being of an inverse
symmetry breaking (ISB).  The opposite case is also possible, where 
the intermediate phase would display a more symmetrical state and then,
later, at higher temperatures, be displaying less symmetry. This would be
an example of what we can call symmetry nonrestoration (SNR). Both ISB and
SNR are also possible states that can emerge in high-energy physics.
Systems exhibiting such forms of phenomena can be constructed in terms
of multiple scalar fields with both self-couplings and intercouplings
and were first investigated by Weinberg in Ref.~\cite{Weinberg:1974hy}
in the context of a two coupled scalar field system at finite
temperature in quantum field theory. 

The type of model investigated in Ref.~\cite{Weinberg:1974hy} has
since then been investigated under different
approaches~\cite{Mohapatra:1979qt,Klimenko:1988mb,Bimonte:1995xs,AmelinoCamelia:1996hw,Orloff:1996yn,Roos:1995vm,Jansen:1998rj,Bimonte:1999tw,Pinto:1999pg}
and provided further motivations for the study of ISB/SNR-like
phenomena for those systems of coupled scalar fields at high
temperatures.  In particular, there has been an increased and renewed
interest in studies involving ISB and SNR  recently, particularly
associated with phase transitions in beyond standard models and
applications~\cite{Meade:2018saz,Baldes:2018nel,Matsedonskyi:2020mlz,Matsedonskyi:2020kuy,Bajc:2020yvd,Chai:2020zgq,Chaudhuri:2020xxb,Chaudhuri:2021dsq,Niemi:2021qvp,Ramazanov:2021eya}.

In this work we are motivated in investigating the interplay of both
temperature and external fields, more specifically, an external
magnetic field, in the phase transition patterns for a system of
coupled complex scalar fields, symmetric under a global $U(1)\times
U(1)$ symmetry.  With the ISB and SNR phenomena gaining recently more interest
in different contexts, the
investigation of not only temperature as a cause of phase transition
in these models,
but also when the system becomes subject to other external
conditions, becomes pertinent. Phase transitions in the presence of external magnetic
fields are relevant for understanding many important physical
systems, ranging from superconductor materials in the context of
condensed matter models~\cite{Maniv:2001zz}, in astrophysical
systems~\cite{Lai:2014nma}, in QCD and heavy-ion collision
problems~\cite{Andersen:2014xxa}, and in the early
Universe~\cite{Grasso:2000wj}. In particular, it is well
known that magnetic fields can lead to phenomena enhancing the symmetry breaking, e.g., the
magnetic catalysis~\cite{Shovkovy:2012zn}, or leading to an opposite effect, the
inverse magnetic catalysis~\cite{Bandyopadhyay:2020zte}. Thus, magnetic
fields can compete with the temperature in nontrivial ways.  The
motivation for the present work is to understand this interplay of
temperature and magnetic field effects as far as the ISB and SNR
phenomena are concerned. To our knowledge, no previous work have
studied such combined effects in the context of ISB and SNR. Hence, in
our opinion, this makes the present study of relevance, in particular
in the context of the aforementioned motivations for studying the
combined effects of temperature and magnetic fields in phase
transitions in complex systems in general. 

To carry out the present study, we will make use of the
nonperturbative method of the optimized perturbation theory
(OPT)~\cite{Stevenson:1981vj,Okopinska:1987hp,Klimenko:1992av,Kleinert:1998zz,Chiku:1998kd,Pinto:1999py,Pinto:1999pg,Farias:2008fs}
(for a recent review on the OPT method, see also
Ref.~\cite{Yukalov:2019nhu} and references therein).  This will help
us to gauge the consistency of the ISB and SNR phenomena against the
perturbative and one-loop expansions when applied to the present problem.  

The remainder of this work is organized as follows. In
Sec.~\ref{sec2}, we introduce the model and its
implementation in the context of the nonperturbative OPT method.  
In Sec.~\ref{sec3}, the effective potential is explicit derived in the OPT method
at first order. The relevant expressions at finite temperature and in an external
magnetic field are given. Some general results for the critical points and phase structure of
the model are given in Sec.~\ref{sec4}. In Sec.~\ref{sec5} we present several numerical results
representative of the effects produced by both temperature and external magnetic field.
{}Finally, our conclusions are given in Sec.~\ref{conclusions}.
One appendix is included to show some of the technical details for the renormalization
of the model. 

\section{The model and its OPT implementation}
\label{sec2}

We consider a model with two coupled complex scalar fields, symmetric under
a global $U(1)\times U(1)$ symmetry.  The most general Lagrangian
density with renormalizable interactions and preserving the global
symmetry is given by
\begin{eqnarray}
\mathcal{L} &=&  ( \partial _{\mu }\phi) (\partial^{\mu }\phi^{*})-
m_{\phi }^{2}(\phi \phi^{*}) -\frac{\lambda _{\phi }}{6} (\phi
\phi^{*})^{2}  \notag \\ &+& ( \partial _{\mu }\psi) (\partial^{\mu
}\psi^{*})  -m_{\psi }^{2}(\psi \psi^{*})  -\frac{\lambda _{\psi }}{6}
(\psi \psi^{*})^{2} \nonumber \\ &-&\lambda
(\phi\phi^{*})(\psi\psi^{*}).
\label{lagrangian}
\end{eqnarray}
The stability of the potential requires the self-couplings
to satisfy $\lambda_\phi > 0$ and $\lambda_\psi > 0$, but
the intercoupling $\lambda$ can be negative, provided that
$\lambda_\phi \lambda_\psi > 9 \lambda^2$.

We work with the model described by Eq.~(\ref{lagrangian}) in the
context of the nonperturbative OPT method.  The implementation of the
OPT  is given by an interpolation of the Lagrangian density in the
form
\begin{equation}
  \mathcal{L} \rightarrow\mathcal{L}^{\delta} =
  (1-\delta)\mathcal{L}_{0}(\eta) + \delta \mathcal{L} , 
\label{interpol}
\end{equation}
where $\mathcal{L}_{0} $ is the Lagrangian density of the free theory,
which is modified by an arbitrary mass parameter $\eta$, while
$\delta$ works as a bookkeeping  (dimensionless)  parameter used only
to keep track of the order at which the OPT is implemented and it is set
to $1$ at the end.
 
Applying the standard interpolation given by Eq.~(\ref{interpol}) to
Eq.~(\ref{lagrangian}) gives
\begin{eqnarray}
\mathcal{L} &=&  ( \partial _{\mu }\phi) (\partial^{\mu }\phi^{*})-
\Omega_{\phi }^{2}(\phi \phi^{*})  -\delta\frac{\lambda _{\phi }}{6}
(\phi \phi^{*})^{2}  \notag \\ &+& ( \partial _{\mu }\psi)
(\partial^{\mu }\psi^{*})  -\Omega_{\psi }^{2}(\psi \psi^{*})
-\delta\frac{\lambda _{\psi }}{6} (\psi \psi^{*})^{2} \nonumber
\\ &-&\delta \lambda (\phi\phi^{*})(\psi\psi^{*}) +\delta \eta_{\phi
}^{2}(\phi \phi^{*}) + \delta\eta_{\psi }^{2}(\psi \psi^{*}),
\label{OPTL}
\end{eqnarray}
where $\Omega _{\phi }^{2}=m_{\phi }^{2}+\eta _{\phi }^{2}$,
$\Omega_{\psi }^{2}=m_{\psi }^{2}+\eta _{\psi }^{2}$, and
$\eta_{\phi,\psi}$ are the OPT (mass) parameters.

As usual in the OPT method, any quantity evaluated at any finite order
in $\delta$ will depend explicitly on the mass parameters 
$\eta_{\phi, \psi}$. These parameters are then fixed by an appropriate
variational principle applied to the physical quantity that is being
computed.  As in most OPT studies, here we make use of the  principle
of minimal sensitivity (PMS)~\cite{Stevenson:1981vj}
\footnote{{}For other ways of fixing the mass parameters of the OPT
  and how they compared to the PMS, see for instance
  Ref.~\cite{Rosa:2016czs}.}. As we are interested in the phase
structure of the model, we also follow the  prescription of previous
works~\cite{Kneur:2007vj,Farias:2008fs,Kneur:2010yv,Duarte:2011ph,Duarte:2017zdz}
of applying the PMS directly to the effective potential computed in
the OPT method. Applying the PMS directly to the effective potential
is also quite convenient, since already at first order in $\delta$ it
is able to produce nontrivial solutions leading to nonperturbative
results. As demonstrated in previous works~\cite{Rosa:2016czs}, this
is also very convenient given the fast convergence of the OPT/PMS 
method\footnote{We point out that the fast convergence of the OPT/PMS
  method has been demonstrated in particular to critical
  three-dimensional models~\cite{Kneur:2002kq}. {}For a recent
  discussion about the convergence properties of the OPT method under
  different optimization procedures, see, e.g.,
  Ref.\cite{Yukalov:2019nhu}.}. Let then $V_{\rm eff} ^{\left(
  k\right) }$ be the effective potential evaluated at order $\delta^k$ in
the OPT. The PMS principle for the present application then reads 
\begin{eqnarray}
&&\frac{d V_{\rm eff} ^{\left( k\right) }}{d\eta_{\phi}}
\Bigr|_{\eta_\phi=\bar{\eta}_{\phi},\eta_\psi=\bar{\eta}_{\psi},\delta
  =1}=0,\;\;\;\; \nonumber\\  
  &&\frac{d V_{\rm eff} ^{\left( k\right)
}}{d\eta_{\psi}}
\Bigr|_{\eta_\phi=\bar{\eta}_{\phi},\eta_\psi=\bar{\eta}_{\psi},\delta
  =1}=0. \label{pms}
\end{eqnarray}
The optimum values $\bar{\eta}_{\phi},\, \bar{\eta}_{\psi}$ are
functions of the original parameters of the theory and it is through
them that the OPT produces nonperturbative results. 

\section{The effective potential for the $U(1)\times U(1)$ model in
  the OPT method}
\label{sec3}

{}From the interpolated Lagrangian density equation (\ref{OPTL}), we can
construct all contributions to the effective potential in the OPT
method.  By  expressing the complex scalar fields $\phi$ and $\psi$ in
terms of their real and imaginary components, $\phi = (\phi_1+i
\phi_2)/\sqrt{2}$ and $\psi = (\psi_1+i \psi_2)/\sqrt{2}$, respectively,
we follow
the usual prescription for obtaining the effective potential. We first
shift the fields around the respective background expectations
values, which can be taken along the real components of the fields
without loss of generality,
$\phi_1 \to \phi_1^\prime = \phi_1 + \phi_0$ and $\psi_1 \to
\psi_1^\prime = \psi_1 + \psi_0$, with $\langle \phi_1 \rangle =
\langle \phi_2 \rangle = \langle \psi_1 \rangle = \langle \psi_1
\rangle = 0$ and  $\langle \phi_1^\prime \rangle = \phi_0$ and
$\langle \psi_1^\prime \rangle = \psi_0$. Thus, at first order in the
OPT, i.e., at first order in $\delta$, the effective potential is
explicitly given by
\begin{eqnarray}
V_{\rm eff} &=& V_{0}(\phi_0,\psi_0)+\sum_{P}\!\!\!\!\!\!\!\!\int \ln
\left(P^{2}+\Omega _{\phi }^{2}\right)  +\sum_{P}\!\!\!\!\!\!\!\!\int
\ln \left(P^{2}+\Omega _{\psi }^{2}\right)  \nonumber \\ &-& \delta
\eta_{\phi }^{2}\sum_{P}\!\!\!\!\!\!\!\!\int \frac{1}{P^{2}+\Omega
  _{\phi }^{2}} -\delta \eta_{\psi}^{2}\sum_{P}\!\!\!\!\!\!\!\!\int
\frac{1}{P^{2}+\Omega _{\psi }^{2}}  \nonumber \\ &+&\delta \frac{
  \lambda _{\phi }}{3}\phi_0^{2}\sum_{P}\!\!\!\!\!\!\!\!\int
\frac{1}{P^{2}+\Omega _{\phi }^{2}} +\delta
\frac{\lambda_\psi}{3}\psi_0^{2}\sum_{P}\!\!\!\!\!\!\!\!\int\frac{1}{P^{2}+\Omega
  _{\psi }^{2}}   \nonumber \\ &+&\delta\frac{\lambda}{2}
\psi_0^{2}\sum_{P}\!\!\!\!\!\!\!\!\int \frac{1}{P^{2}+\Omega _{\phi
  }^{2}} +\delta
\frac{\lambda}{2}\phi_{0}^{2}\sum_{P}\!\!\!\!\!\!\!\!\int
\frac{1}{P^{2}+\Omega _{\psi }^{2}} \nonumber \\ &+&\delta
\frac{\lambda_\phi}{3}\left[
  \sum_{P}\!\!\!\!\!\!\!\!\int\frac{1}{P^{2}+\Omega _{\phi
    }^{2}}\right] ^{2} +\delta\frac{\lambda_\psi}{3}\left[
  \sum_{P}\!\!\!\!\!\!\!\!\int \frac{1}{P^{2}+\Omega
    _{\psi}^{2}}\right] ^{2} \nonumber \\ &+&\delta \lambda\left[
  \sum_{P}\!\!\!\!\!\!\!\!\int \frac{1}{P^{2}+\Omega_{\phi
    }^{2}}\right]  \left[ \sum_{P}\!\!\!\!\!\!\!\!\int
  \frac{1}{P^{2}+\Omega _{\psi }^{2}}\right]    \nonumber
\\ &+&\frac{1}{16 \pi^2 \epsilon}\left( \frac{2 \delta
  \lambda_\phi}{3}\Omega_\phi^2 +  \delta \lambda \Omega_\psi^2
\right)\sum_{P}\!\!\!\!\!\!\!\!\int \frac{1}{P^{2}+\Omega _{\phi
  }^{2}} \nonumber \\ &+&\frac{1}{16 \pi^2 \epsilon}\left( \frac{2
  \delta \lambda_\psi}{3}\Omega_\psi^2 +  \delta \lambda \Omega_\phi^2
\right)\sum_{P}\!\!\!\!\!\!\!\!\int \frac{1}{P^{2}+\Omega _{\psi
  }^{2}}, \nonumber \\
\label{fullVeff}
\end{eqnarray}
where the last two terms shown in Eq.~(\ref{fullVeff}), proportional
to $1/\epsilon$, are the contributions at order $\delta$ generated by
the mass counterterms needed for renormalization (note that throughout
this paper, we work with dimensional regularization in the
$\overline{\mathrm{MS}}$ scheme).  The tree-level contribution in
Eq.~(\ref{fullVeff}), $V_{0}(\phi_0,\psi_0)$, is given by
\begin{eqnarray}
 V_{0}(\phi_0,\psi_0) &=&\frac{1}{2}\Omega _{\phi
 }^{2}\phi_0^{2}+\frac{1}{2}\Omega _{\psi
 }^{2}\psi_0^{2}-\frac{1}{2}\delta \eta _{\phi }^{2}\phi_0^{2}-
 \frac{1}{2}\delta \eta _{\psi }^{2}\psi_0^{2} \nonumber
 \\ &+&\frac{1}{24}\delta \lambda _{\phi
 }\phi_0^{4}+\frac{1}{24}\delta \lambda _{\psi
 }\psi_0^{4}+\frac{1}{4}\delta \lambda \phi_0^{2}\psi _{0}^{2}
 \nonumber \\ &+& \frac{1}{2}\Delta m_\phi \phi_0^2 +\frac{1}{2}\Delta
 m_\psi \psi_0^2 +   \Delta V, \label{Fzero}
\end{eqnarray}
with the required counterterms at ${\cal O}(\delta)$ in the OPT scheme
given by the terms shown in the last line in Eq.~(\ref{Fzero}). Note that
at the present order of the OPT, one only requires the mass
counterterms $ \Delta m_\phi$ and $\Delta m_\psi $, for the $\phi$ and
$\psi$ fields, respectively, and a vacuum counterterm $\Delta V$
(vertice counterterms are required when carrying out the OPT at second
order).
Details of the renormalization scheme used here to renormalize
Eq.~(\ref{fullVeff}) are given in the Appendix.
{}Furthermore, in the notation used in Eq.~(\ref{fullVeff}) for the
momentum integrals, which are expressed in Euclidean spacetime, we
have in the finite-temperature case that
\begin{equation}
\sum_{P}\!\!\!\!\!\!\!\!\int \equiv T\sum_{P_{4}=\omega_n}\left(
\frac{e^{\gamma_E }M^{2}}{4\pi } \right) ^{\epsilon }\int
\frac{d^{d}p}{\left( 2\pi \right) ^{d}},
\label{sumT}
\end{equation}
with divergent vacuum momentum integral terms regularized in the
$\overline{\mathrm{MS}}$ scheme, with $d=3-2\epsilon $, $\gamma_{E}$
is the Euler-Mascheroni constant and $M$ is the arbitrary mass
regularization scale. The sum in Eq.~(\ref{sumT}) is performed over
the Matsubara frequencies for bosons, $\omega_n= 2\pi n T$, $n\in
\mathbb{Z}$.  The field propagators in Eq.~(\ref{fullVeff}) are
such that $P^{2}+\Omega^{2} \equiv \omega_n^2 + E^2({\bf p})$, with
dispersion relation
\begin{equation}
E^2({\bf p})={\bf p}^2+\Omega^2.
\label{Ep}
\end{equation}
Likewise, when the external magnetic field $B$ is coupled to the
system, we now have instead for the regularized momentum integrals
that
\begin{equation}
\sum_{P}\!\!\!\!\!\!\!\!\int \equiv  \frac{ eB }{2\pi }
\sum_{k=0}^{+\infty } T \sum_{P_{4}=\omega_n}  \left(
\frac{e^{\gamma_E }M^{2}}{4\pi }\right) ^{\epsilon
}\int\frac{d^{d-2}p_z}{\left( 2\pi\right) ^{d-2}},
\label{sumTB}
\end{equation}
where the first sum in Eq.~(\ref{sumTB}) is over the Landau levels,
with $k\in \mathbb{N}$ and $e$ is the elementary charge. Here, the
dispersion relation for the bosons is
\begin{equation}
E^2({\bf p}) \to E^{2}\left( p_{z},k\right)
=p_{z}^{2}+\Omega^{2}+\left( 2k+1\right)   eB.
\label{EpB}
\end{equation}
Note also that in writing Eqs.~(\ref{sumTB}) and (\ref{EpB}), without
loss of generality, a gauge was chosen such that the magnetic field is
in the $z$ direction.

\subsection{The renormalized effective potential at ${\cal O}(\delta)$}

After performing the Matsubara sums, we obtain for the renormalized
effective potential the result at ${\cal O}(\delta)$,
\begin{eqnarray}
V_{\rm eff,R}^{(\delta)} &=&  \frac{1}{2}\Omega _{\phi
}^{2}\phi_0^{2}+\frac{1}{2}\Omega _{\psi
}^{2}\psi_0^{2}-\frac{1}{2}\delta \eta _{\phi }^{2}\phi_0^{2}-
\frac{1}{2}\delta \eta _{\psi }^{2}\psi_0^{2} \nonumber
\\ &+&\frac{1}{4!}\delta \lambda _{\phi }\phi_0^{4}+\frac{1}{4!}\delta
\lambda _{\psi }\psi_0^{4}+\frac{1}{4}\delta \lambda
\phi_0^{2}\psi_{0}^{2}  \nonumber \\ &+& Y_\phi+ Y_\psi -\delta \eta
_{\phi }^{2} X_\phi  -\delta \eta _{\psi }^{2} X_\psi \nonumber \\ &+&
\frac{1}{3}\delta \lambda _{\phi }\phi_0^{2} X_\phi +
\frac{1}{3}\delta \lambda _{\psi }\psi_0^{2} X_\psi +
\frac{1}{2}\delta \lambda \psi_0^{2} X_\phi + \frac{1}{2}\delta
\lambda \phi_0^{2}X_\psi  \nonumber \\ &+& \frac{1}{3}\delta \lambda
_{\phi}X^2_\phi +\frac{1}{3}\delta \lambda _{\psi}X^2_\psi +\delta
\lambda X_\phi X_\psi,
\label{VeffR}
\end{eqnarray}
where we are using a simplified notation in
Eq.~(\ref{VeffR}) for the $Y_{\phi,\psi}$ and $X_{\phi,\psi}$ terms,
which will depend on the three cases that we will be studying here: (a) at
$T\neq 0$ and $B=0$; (b) at $T\neq 0$ and $B\neq 0$; and (c) $T= 0$ and
$B\neq 0$. 

\subsubsection{Case (a): $T\neq 0$ and $B=0$}
\label{casea}

At finite temperature and in the absence of an external magnetic field
($B=0$), we have that
\begin{equation}
Y\equiv Y(T) = - \frac{1}{2 (4 \pi)^2} \left[ \frac{3}{2} +
  \ln\left(\frac{M^2}{\Omega^2}\right) \right] \Omega^4 +J(T),
\label{YT}
\end{equation}
where
\begin{eqnarray}
J(T ) &=&   \frac{T^4}{\pi^2} \int_0^\infty dz \, z^2
\ln\left[1-\exp\left(-\sqrt{z^2+\Omega^2/T^2}\right)\right],
\nonumber \\
\label{JB}
\end{eqnarray}
and
\begin{equation}
X \equiv X(T) = \frac{\Omega^{2}}{16\pi ^{2}}\left[ \ln \left(
  \frac{\Omega^{2}}{ M^{2}}\right) -1\right] +I(T),
\label{XT}
\end{equation}
where
\begin{eqnarray}
I(T) &=& \frac{T^2}{2 \pi^2}\int_0^\infty dz
\frac{z^2}{\sqrt{z^2+\frac{\Omega^2}{T^2}}}\;\frac{1}{
  \exp\left(\sqrt{z^2+\frac{\Omega^2}{T^2}}\right)-1}.  \nonumber \\
\label{IB}
\end{eqnarray}

\subsubsection{Case (b): $T\neq 0$ and $B\neq 0$}
\label{caseb}

At finite temperature and in the presence of an external magnetic
field ($B\neq 0$), we have that
\begin{eqnarray}
Y\equiv Y(B,T) &=& -\frac{\Omega^4}{32 \pi^2}\left[ 1 +\ln \left(
  \frac{M^2}{2 eB}\right)\right]   \nonumber \\ &+& \frac{(eB)^2}{4
  \pi^2} \zeta'\left(-1,\frac{\Omega^2 + eB}{2 eB}\right) \nonumber
\\ &+& J(B,T),
\label{YBT}
\end{eqnarray}
where  $\zeta'(s,a)$ is the $s$ derivative of the Hurwitz zeta
function~\cite{Elizalde:1994gf},
\begin{equation}
\zeta(s,a) = \sum_{k=0}^{\infty} \frac{1}{(k+a)^s}\;,
\label{zetafunc}
\end{equation}
and 
\begin{eqnarray}
J(B,T ) &=&  \frac{e B}{\pi} T^2 \sum_{k=0}^{+\infty}
\int_{-\infty}^{+\infty} \frac{d z}{2 \pi}  \nonumber \\ &\times & \ln
\left\{ 1- \exp\left[ - \sqrt{z^2 + \frac{\Omega^2}{T^2} +  (2k +
    1)\frac{e B}{T^2}} \right] \right\},  \nonumber \\
\label{JBTB}
\end{eqnarray}
while for $X$ we now have that
\begin{eqnarray}
X \equiv X(B,T) &=& \frac{e B}{8 \pi^2} \ln \left[ \Gamma \left(
  \frac{ \Omega^2 + e B}{2 e B} \right) \right]  \nonumber
\\ &-&\frac{e B}{16 \pi^2} \ln(2 \pi) - \frac{\Omega^2}{16 \pi^2}
\ln\left( \frac{M^2}{2 e B} \right)\nonumber \\ &+& I(B,T),
\label{XBT}
\end{eqnarray}
where
\begin{eqnarray}
I(B,T) &=& \frac{e B}{2 \pi} \sum_{k=0}^{+\infty}
\int_{-\infty}^{+\infty} \frac{d z}{2 \pi} \frac{1}{\sqrt{z^2 +
    \frac{\Omega^2}{T^2} +  (2k + 1)\frac{e B}{T^2}}} \nonumber
\\ &\times & \frac{1}{\exp\left[\sqrt{z^2 + \frac{\Omega^2}{T^2} +
      (2k + 1)\frac{e B}{T^2}}\right] -1}.
\label{IBTB}
\end{eqnarray}

\subsubsection{Case (c): $T= 0$ and $B\neq 0$}
\label{casec}

{}Finally, at zero temperature ($T=0$) but in the presence of the
external magnetic field, we have that in the above expressions $J(B,T
)=0$ and $ I(B,T)=0$ in Eqs.~(\ref{YBT}) and (\ref{XBT}),
respectively. Thus,
\begin{eqnarray}
Y(B,T=0) &=& -\frac{\Omega^4}{32 \pi^2}\left[ 1 +\ln \left(
  \frac{M^2}{2 eB}\right)\right]   \nonumber \\ &+& \frac{(eB)^2}{4
  \pi^2} \zeta'\left(-1,\frac{\Omega^2 + eB}{2 eB}\right),
\label{YBT0}
\end{eqnarray}
and 
\begin{eqnarray}
\!\!\!\!\!\!\!\!X(B,T=0) &=& \frac{e B}{8 \pi^2} \ln \left[ \Gamma
  \left( \frac{ \Omega^2 + e B}{2 e B} \right) \right]  \nonumber
\\ &-&\frac{e B}{16 \pi^2} \ln(2 \pi) - \frac{\Omega^2}{16 \pi^2}
\ln\left( \frac{M^2}{2 e B} \right).
\label{XBT0}
\end{eqnarray}

\subsection{PMS and background expectation values for the fields}

Applying the PMS procedure Eq.~(\ref{pms}) to the renormalized
effective potential Eq.~(\ref{VeffR}), we obtain that
$\bar{\eta}_\phi$ and $\bar{\eta}_\psi$ are obtained from the coupled
equations,
\begin{equation}
\bar{\eta}_\phi^{2}=\frac{\lambda_\phi
}{3}\tilde{\phi}^2+\frac{\lambda}{2} \tilde{\psi}^2
+\frac{2\lambda_\phi }{3} X_\phi\Bigr|_{\eta
  _\phi=\bar{\eta}_\phi}+\lambda X_\psi\Bigr|_{\eta_\psi
  =\bar{\eta}_\psi}, \label{pmsetaphi}
\end{equation}
\begin{equation}
\bar{\eta}_\psi^2=\frac{\lambda_\psi}{3}\tilde{\psi}^2+\frac{\lambda}{2}
\tilde{\phi}^2 +\frac{2\lambda_\psi }{3} X_\psi\Bigr|_{\eta_\psi
  =\bar{\eta}_\psi}+\lambda X_\phi\Bigr|_{\eta_\phi=\bar{\eta}_\phi
}, \label{pmsetapsi}
\end{equation}
which are to be solved together with the ones defining the background
field values  $\tilde{\phi}$ and $\tilde{\psi}$, obtained, respectively, from
\begin{equation}
\frac{\partial V_{\rm eff, R}}{\partial \phi_0}\Bigr|_{\phi_0=\tilde
  \phi, \psi_0=\tilde \psi}=0, \;\;\;\; \frac{\partial V_{\rm eff,
    R}}{\partial \psi_0}\Bigr|_{\phi_0=\tilde \phi, \psi_0=\tilde
  \psi}=0,
\end{equation}
which lead to the trivial solutions $\tilde \phi=\tilde \psi=0$, and
the other two nontrivial solutions, corresponding to the gap
equations, given, respectively, by
\begin{equation}
m_{\phi }^{2}+\frac{\lambda _{\phi
}}{6}\tilde{\phi}^{2}+\frac{\lambda}{2}
\tilde{\psi}^{2}+\frac{2\lambda _{\phi }}{3} X_\phi
\Bigr|_{\eta_\phi=\bar{\eta}_\phi} +\lambda X_\psi\Bigr|_{\eta_\psi
  =\bar{\eta}_\psi} =0, \label{VEVphi}
\end{equation}
and
\begin{equation}
m_{\psi }^{2}+\frac{\lambda _{\psi
}}{6}\tilde{\psi}^{2}+\frac{\lambda}{2}
\tilde{\phi}^{2}+\frac{2\lambda_{\psi }}{3}X_\psi
\Bigr|_{\eta_\psi=\bar{\eta}_\psi} +\lambda X_\phi\Bigr|_{\eta_\phi
  =\bar{\eta}_\phi} =0. \label{VEVpsi}
\end{equation}

We can define the effective square masses of the fields
which will include all effects of $T$ and $B$ at  ${\cal O}(\delta)$ in the
OPT-PMS. {}From the second derivative in the background fields
of the effective potential Eq.~(\ref{VeffR}), we obtain
\begin{eqnarray}
&& \!\!\!\!\!\!\!\!\!\!\!\!\! m_{\phi,{\rm eff}}^2 \!=\! m_\phi^2
  +\frac{\lambda}{2}\tilde{\psi}^2+ \frac{2 \lambda_\phi}{3}
  X_\phi\Bigr|_{\eta _\phi=\bar{\eta}_\phi} + \lambda
  X_\psi\Bigr|_{\eta _\psi=\bar{\eta}_\psi},
\label{mphieff}
\\ && \!\!\!\!\!\!\!\!\!\!\!\!\!  m_{\psi,{\rm eff}}^2 \!=\! m_\psi^2
+\frac{\lambda}{2}\tilde{\phi}^2+ \frac{2 \lambda_\psi}{3}
X_\psi\Bigr|_{\eta _\psi=\bar{\eta}_\psi} + \lambda X_\phi\Bigr|_{\eta
  _\phi=\bar{\eta}_\phi}.
\label{mpsieff} 
\end{eqnarray}

{}Finally, note that Eqs.~(\ref{mphieff}) and (\ref{mpsieff}), which
give the curvature of the effective potential Eq.~(\ref{VeffR}) in the
$\phi$ and $\psi$ directions, respectively, should not be confused
with the {\it physical masses} for the fields. {}For instance, the real and imaginary
components for the $\phi$ and $\psi$ fields, $\phi_1, \; \phi_2$, for
the complex scalar field $\phi$, and  $\psi_1, \; \psi_2$, for the
complex scalar field $\psi$, will have effective physical masses at
${\cal O}(\delta)$ given, respectively, by
\begin{eqnarray}
m_{\phi_1,{\rm eff}}^2 &=& m_\phi^2 +
\frac{\lambda_\phi}{2}\tilde{\phi}^2 +\frac{\lambda}{2}\tilde{\psi}^2
\nonumber \\ &+& \frac{2 \lambda_\phi}{3} X_\phi\Bigr|_{\eta
  _\phi=\bar{\eta}_\phi} +  \lambda X_\psi\Bigr|_{\eta
  _\psi=\bar{\eta}_\psi},
\label{mphi1eff}
\\ m_{\phi_2,{\rm eff}}^2 &=& m_\phi^2 +
\frac{\lambda_\phi}{6}\tilde{\phi}^2 +\frac{\lambda}{2}\tilde{\psi}^2
\nonumber \\ &+& \frac{2 \lambda_\phi}{3} X_\phi\Bigr|_{\eta
  _\phi=\bar{\eta}_\phi} +  \lambda X_\psi\Bigr|_{\eta
  _\psi=\bar{\eta}_\psi},
\label{mphi2eff}
\\ m_{\psi_1,{\rm eff}}^2 &=& m_\psi^2
+\frac{\lambda_\psi}{2}\tilde{\psi}^2 +
\frac{\lambda}{2}\tilde{\phi}^2 \nonumber \\ &+& \frac{2
  \lambda_\psi}{3} X_\psi\Bigr|_{\eta _\psi=\bar{\eta}_\psi} +
\lambda X_\phi\Bigr|_{\eta _\phi=\bar{\eta}_\phi},
\label{mpsi1eff} 
\\ m_{\psi_2,{\rm eff}}^2 &=& m_\psi^2
+\frac{\lambda_\psi}{6}\tilde{\psi}^2 +
\frac{\lambda}{2}\tilde{\phi}^2 \nonumber \\ &+& \frac{2
  \lambda_\psi}{3} X_\psi\Bigr|_{\eta _\psi=\bar{\eta}_\psi} +
\lambda X_\phi\Bigr|_{\eta _\phi=\bar{\eta}_\phi}.
\label{mpsi2eff} 
\end{eqnarray}
With our choice of shifting the fields around their real components,
we have that $m_{\phi_1,{\rm eff}}$ and $m_{\psi_1,{\rm eff}}$
correspond to the Higgs modes for the $\phi$ and $\psi$ fields,
respectively, while  $m_{\phi_2,{\rm eff}}$ and $m_{\psi_2,{\rm eff}}$
correspond to the Goldstone modes for the fields.  Note that these
masses remain positive definite always. In particular, from the
previous equations we can easily demonstrate the validity of the
Goldstone theorem in the OPT method (see, in particular,
Ref.~\cite{Duarte:2011ph} for an explicit proof in the case of one
complex scalar field. An analogous demonstration also follows here).

Equations (\ref{pmsetaphi}), (\ref{pmsetapsi}), (\ref{VEVphi}),
(\ref{VEVpsi}), (\ref{mphieff}), and (\ref{mpsieff}) allow us to make
the complete analysis of the possible transition patterns at finite
$T$ and/or $B$ that are possible for the model.

\section{General results for the critical points and phase transition patterns}
\label{sec4}

By combining Eqs.~(\ref{pmsetaphi}), (\ref{pmsetapsi}),
(\ref{mphieff}), and (\ref{mpsieff}), we straightforwardly find that
\begin{eqnarray}
&& m_\phi^2 + \bar{\eta}_\phi^2\equiv \Omega_\phi^2 =
  \frac{\lambda_\phi}{3} \tilde{\phi}^2 + m_{\phi,{\rm eff}}^2,
\label{Omegaphi}
\\ && m_\psi^2 + \bar{\eta}_\psi^2\equiv \Omega_\psi^2 =
\frac{\lambda_\psi}{3} \tilde{\psi}^2 + m_{\psi,{\rm eff}}^2,
\label{Omegapsi}
\end{eqnarray}
which are quite general results obtained from the application of the
OPT-PMS at ${\cal O}(\delta)$.

\subsection{Some simple OPT-PMS results at ${\cal O}(\delta)$}

By considering the case (a), $T\neq 0,\, B=0$, from Sec.~\ref{casea},
we have that whenever the system is at the critical point where
$\tilde{\phi}(T_{c,\phi})=0$ and $m_{\phi,{\rm eff}}^2(T_{c,\phi})=0$,
then $\Omega_\phi \to 0$. Likewise, in the case of symmetry
restoration in the direction of $\psi$, where
$\tilde{\psi}(T_{c,\psi})=0$ and $m_{\psi,{\rm eff}}^2(T_{c,\psi})=0$,
then $\Omega_\psi \to 0$. In each of these cases, from Eq.~(\ref{XT})
we find that
\begin{equation} 
X(T=T_c,B=0)\Bigr|_{\Omega =0} = T_c^2/12.
\label{XTc}
\end{equation}
Analogously to the above situation, in the case (c), where $T=0,\, B\neq 0$, from
Sec.~\ref{casec}, at the critical point for symmetry restoration in
the  direction of $\phi$, where $\tilde{\phi}(B_{c,\phi})=0$ and
$m_{\phi,{\rm eff}}^2(B_{c,\phi})=0$,  with
$\tilde{\psi}(B_{c,\phi})=0$, then $\Omega_\phi \to 0$, while in the
case of symmetry restoration in the direction of $\psi$, where
$\tilde{\psi}(B_{c,\psi})=0$ and $m_{\psi,{\rm eff}}^2(B_{c,\psi})=0$,
whenever $\tilde{\phi}(B_{c,\psi})=0$, then $\Omega_\psi \to
0$. Hence, from Eq.~(\ref{XBT0}), we get now that
\begin{equation}
X(T=0,B=B_c)\Bigr|_{\Omega =0} = - e B_c \ln(2)/(4\pi)^2.
\label{XBc}
\end{equation} 
Note that due to the difference in sign for the function $X$ in each
of the above cases, we will have opposite effects on the symmetry behavior of the
system, whenever thermal effects or the external magnetic field
dominates.

It is also useful to combine Eqs.~(\ref{VEVphi}) and (\ref{VEVpsi})
and also use Eqs.~(\ref{mphieff}) and (\ref{mpsieff}) to arrive at
\begin{eqnarray}
\tilde{\phi}^2 =  \left\{
\begin{array}{ll}
-\frac{6m_{\phi,{\rm eff}}^2 }{\lambda_\phi}, & {\rm for}\;\;
m_{\phi,{\rm eff}}^2<0, \\ 0, &{\rm for}\; m_{\phi,{\rm eff}}^2>0,
\end{array}
\right.
\label{phitilde}
\end{eqnarray}
and
\begin{eqnarray}
\tilde{\psi}^2 =  \left\{
\begin{array}{ll}
-\frac{6m_{\psi,{\rm eff}}^2 }{\lambda_\psi}, & {\rm for}\;\;
m_{\psi,{\rm eff}}^2<0, \\ 0, &{\rm for}\; m_{\psi,{\rm eff}}^2>0,
\end{array}
\right.
\label{psitilde}
\end{eqnarray}
where in the effective masses, Eqs.~(\ref{mphieff}) and
(\ref{mpsieff}), $X_\phi$ and $X_\psi$ are evaluated at
$\eta_\phi=\bar{\eta}_{\phi}$ and at $\eta_\psi=\bar{\eta}_{\psi}$,
respectively [i.e., with $\Omega_\phi$ and $\Omega_\psi$ as given by
Eqs.~(\ref{Omegaphi}) and (\ref{Omegapsi}) at ${\cal O}(\delta)$]. 

{}From the above expressions, we can for example set the coupling
between $\phi$ and $\psi$ to zero, $\lambda=0$.  Let us consider for
example the case (a), with $T\neq 0,\, B=0$.  Assuming symmetry
restoration in the directions of the $\phi$ and $\psi$ fields 
happening at $T_{c,\phi}$ and at
$T_{c,\psi}$,  where $\tilde{\phi}(T_{c,\phi})=0$ and
$\tilde{\psi}(T_{c,\psi})=0$, respectively,  we get from, e.g.,
Eq.~(\ref{Omegaphi}) that $\Omega_\phi^2(T=T_{c,\phi}) = m_{\phi,{\rm
    eff}}^2(T=T_{c,\phi})=0$.  Then, from Eq.~(\ref{XTc}), we get,
also using Eq.~(\ref{mphieff}), that $T_{c,\phi}^2 = - 18
m_\phi^2/\lambda_\phi$, a result first derived in
Ref.~\cite{Duarte:2011ph}.  Analogously, using  Eq.~(\ref{mpsieff}),
we have that $T_{c,\psi}^2 = - 18 m_\psi^2/\lambda_\psi$.  It is
important to notice that these results are exact at ${\cal O}(\delta)$
as a consequence of Eqs.~(\ref{Omegaphi}), (\ref{Omegapsi}), and~(\ref{XTc}) and not a result of any high-temperature expansion
at the one-loop or perturbative levels. 

Similarly to the case above, where we set the coupling between $\phi$
and $\psi$ to zero, $\lambda=0$, we can make an analogous analysis for
the case (c), with $T= 0,\, B\neq 0$.  In this case, symmetry
restoration in the $\phi$ and $\psi$ directions happening at $B_{c,\phi}$ and at
$B_{c,\psi}$,  where $\tilde{\phi}(T_{c,\phi})=0$ and
$\tilde{\psi}(T_{c,\psi})=0$, respectively.  We now get from
Eq.~(\ref{Omegaphi}) that $\Omega_\phi(B=B_{c,\phi}) = m_{\phi,{\rm
    eff}}^2(B=B_{c,\phi})=0$.  Then, from Eq.~(\ref{XBc}), we now get
that $eB_{c,\phi} = 24 \pi^2 m_\phi^2/(\lambda_\phi \ln 2)$.
Analogously, using  Eq.~(\ref{mpsieff}), we have that $eB_{c,\psi} =
24 \pi^2 m_\psi^2/(\lambda_\psi \ln 2)$.  While in the case (a)
symmetry is restored at some finite temperature when the system is
initially in the broken phase ($m^2_{\phi,\psi} <0$) at $T=0$, we have
the opposite situation in an external magnetic field. 
Starting with the system in the symmetric phase ($m^2_{\phi,\psi}>0$) at $B=0$, it
will go to a symmetry broken  phase at a sufficiently large external
magnetic field. This is reminiscent of the magnetic catalysis
effect~~\cite{Shovkovy:2012zn}.

When the intercoupling is nonvanishing, $\lambda \neq 0$, the situation gets more
involved. The proper analysis in this case can only be studied
numerically in the present study of the OPT-PMS. However, we can still
get some useful results when analyzing the model in the usual
perturbation theory and using either a high-temperature expansion,
$m_{\phi,\psi}/T \ll 1$, or a high magnetic field approximation,
$m_{\phi,\psi}^2/(eB) \ll 1$. This is useful for later comparison with
the OPT results. 

\subsection{Perturbation theory results}

It is useful to compare the results to be obtained with the OPT for
the model studied here with those obtained using perturbation theory
in the high-temperature and/or high magnetic field case.  In this
case, the expressions for the effective masses Eqs.~(\ref{mphieff})
and (\ref{mpsieff}) still hold and we can also use
Eqs.~(\ref{phitilde}) and (\ref{psitilde}), where we set  $\eta_\phi =
\eta_\psi=0$ in those expressions whenever $m_\phi^2> 0$ and
$m_\psi^2 >0$, i.e., when we start in the symmetry restored phase in both
$\phi$ and $\psi$ directions.  The case of starting in the symmetry
broken phase for both $\phi$ and $\psi$, i.e.,  $m_\phi^2 <0$ and
$m_\psi^2 < 0$, is worked out more conveniently within the one-loop
approximation, instead of using perturbation theory, and it is given
in the next subsection below.  In the high-temperature approximation, 
$m_{\phi,\psi}/T \ll 1$, the thermal integral contributing to the 
effective masses can be approximated at leading order 
as~\cite{Kapusta:2006pm}
\begin{equation} 
X_{\phi,\psi}(T,B=0)\Bigr|_{m_{\phi,\psi}/T \ll 1} \simeq T^2/12.
\label{XhighT}
\end{equation}
Similarly, in the case of a finite magnetic field, using
Eq.~(\ref{XBT0}), we get in the high magnetic field approximation that
\begin{equation}
X_{\phi,\psi}(T=0,B)\Bigr|_{m_{\phi,\psi}^2/(eB) \ll 1} \simeq - e B
\ln(2)/(4\pi)^2.
\label{XhighB}
\end{equation} 
Thus, from Eqs.~(\ref{mphieff}) and (\ref{mpsieff}) in the high-temperature approximation and in the absence of magnetic field, we get
\begin{eqnarray}
&& \!\!\!\!\!\!\!\!\!\! m_{\phi,{\rm eff}}^2\Bigr|_{m_{\phi,\psi}/T
    \ll 1} \simeq m_\phi^2 + \frac{\lambda}{2} \tilde{\psi}^2+
  \left(\frac{2 \lambda_\phi}{3}  + \lambda \right) \frac{T^2}{12},
\label{mphieffhighT}
\\ && \!\!\!\!\!\!\!\!\!\! m_{\psi,{\rm eff}}^2\Bigr|_{m_{\phi,\psi}/T
  \ll 1} \simeq  m_\psi^2 + \frac{\lambda}{2} \tilde{\phi}^2+
\left(\frac{2 \lambda_\psi}{3} + \lambda \right) \frac{T^2}{12}.
\label{mpsieffhighT} 
\end{eqnarray}
When $\lambda <0$ and choosing coupling constants such that, e.g., $2
\lambda_\phi/3  + \lambda <0$, because of the boundedness
condition $\lambda_\phi \lambda_\psi > 9 \lambda^2$, it necessarily  
implies that $2\lambda_\psi/3  + \lambda > 0$. Hence, when $m_\phi^2 > 0$, i.e.,
initially choosing the system to be in the symmetry restored phase in
the $\phi$ direction, Eq.~(\ref{mphieffhighT}) will imply that there
will be an ISB at the approximated critical temperature 
\begin{equation}
T_{c,\phi}^2 \simeq  -\frac{12 m_\phi^2}{2 \lambda_\phi/3  + \lambda},
\label{TcphiPT}
\end{equation}
where we have considered that $\tilde{\psi}(T=T_{c,\phi})=0$,
i.e., the symmetry in the $\psi$ direction is restored at
$T_{c,\psi} < T_{c,\phi}$, when $m_\psi^2<0$, and remains restored,
or when $m_\psi^2> 0$, in which case we always have $\tilde{\psi}=0$.  
In the opposite case, when $m_\phi^2 < 0$, i.e.,
initially choosing the system to be in the symmetry broken phase in
the $\phi$ direction, then it will remain in this state  at
arbitrarily large temperatures\footnote{Arbitrarily large temperatures
  here mean up to those temperatures where the model can be considered
  valid, which should correspond to temperatures below some scale
  $\Lambda_{\rm UV}$. {}For temperatures above $\Lambda_{\rm UV}$ 
the model might require an
  ultraviolet (UV) completion and ISB- and SNR-like phenomena are not
  guarantee to happen~\cite{Bajc:2020yvd}. }. This is the case of SNR.
In the case of considering a finite magnetic field, $B\neq 0$, but 
at $T=0$, the situation is
similar, although the roles of ISB and SNR get reversed because of the
minus sign in Eq.~(\ref{XhighB}).  The situation is analogous when
choosing that there is a symmetry restoration in the $\psi$ direction,
i.e., considering that $m_\phi^2, \, m_\psi^2>0$, with
$2\lambda_\psi/3  + \lambda < 0$, which now implies that 
$2 \lambda_\phi/3  + \lambda > 0$.
Hence, the symmetry remains restored in the $\phi$ field direction at high temperatures,
but in the $\psi$ field direction there is ISB with the critical temperature for $\psi$ 
given by
\begin{equation}
T_{c,\psi}^2 \simeq  -\frac{12 m_\psi^2}{2 \lambda_\psi/3  + \lambda}.
\label{TcpsiPT}
\end{equation}

As already discussed before, in the case of a finite magnetic field
and at zero temperature, because of the minus sign in
Eq.~(\ref{XhighB}) the situation gets reversed with respect to what it
is obtained in the finite-temperature case. In the case of
perturbation theory at large magnetic fields, we can extend the
results in Eqs.~(\ref{mphieffhighT}) and (\ref{mpsieffhighT}) by
making the replacement $T^2/12 \to -eB \ln(2)/(4 \pi)^2$.  Hence,
starting in the symmetry restored phase for both the $\phi$- and $\psi$-
field directions, i.e., considering 
$m_\phi^2> 0$ and $m_\psi^2 >0$, and assuming $2 \lambda_\phi/3 +
\lambda <0$ (with $\lambda <0$), we have that the symmetry will tend
to remain restored in the $\phi$ direction, while in the
$\psi$ direction there is ISB at a critical magnetic field, 
\begin{equation}
eB_{c,\psi} \simeq \frac{(4\pi)^2 m_\psi^2/\ln(2)}{2 \lambda_\psi/3  +
  \lambda}.
\label{BcpsiPT}
\end{equation}
In the case of starting in the symmetry broken phase for both field
directions, $m_\phi^2< 0$ and $m_\psi^2 <0$, and assuming again that
$2 \lambda_\phi/3 + \lambda <0$ the situation gets more involved since
the results also depend on how the background fields $\tilde \phi$
and $\tilde \psi$ will behave at large $B$.  As we are going to see
numerically in Sec.~\ref{sec5B}, the tendency is both $\phi$ and
$\psi$ to remain  in a symmetry broken state.

\subsection{One-loop effective potential}

In the case of starting in the symmetry broken phase for both $\phi$
and $\psi$, i.e.,  $m_\phi^2 <0$ and $m_\psi^2 < 0$, it is more
convenient to work with the one-loop effective potential.  The
one-loop effective potential for the background fields $\phi_0$ and
$\psi_0$ given by the general expression
\begin{eqnarray}
V_{\rm eff}&=& \frac{m_{\phi }^{2}}{2}\phi_0^{2}+\frac{m_{\psi
  }^{2}}{2}\psi_0^{2}+\frac{\lambda_\phi}{4!}\phi_0^{4}
+\frac{\lambda_\psi}{4!}\psi_0^{4}+\frac{\lambda}{4}
\phi_0^{2}\psi_0^{2} \nonumber \\ &+&\frac{1}{2}\sum_i
\sum_{P}\!\!\!\!\!\!\!\!\int \ln
\left[P^2+\mathcal{M}_i^2(\phi_0,\psi_0)\right],
\label{Veff1loop}
\end{eqnarray}
where $\mathcal{M}_i^2(\phi_0,\psi_0)$ denotes the physical masses
eigenstates of the model, i.e., expressing the mass eigenvalues in the
Higgs and Goldstone basis. These are explicitly given by the
diagonalization of the quadratic mass matrix for the fields,
\begin{widetext}
\begin{equation}
\hat{\mathcal{M}}(\phi_0,\psi_0)= \left(
\begin{array}{cccc}
 m_{\phi }^2+\frac{\lambda _{\phi } \phi_0^2}{2} +\frac{\lambda  \psi
   _0^2}{2} & 0 & \lambda  \phi_0 \psi_0 & 0 \\ 0 & m_{\phi
 }^2+\frac{\lambda _{\phi } \phi_0^2}{6} +\frac{\lambda  \psi _0^2}{2}
 & 0 & 0 \\ \lambda  \phi_0 \psi_0 & 0 & m_{\psi }^2+\frac{\lambda
   \phi _0^2}{2}+\frac{\lambda _{\psi } \psi_0^2}{2}  & 0 \\ 0 & 0 & 0
 & m_{\psi }^2+\frac{\lambda  \phi_0^2}{2}+\frac{\lambda _{\psi }
   \psi_0^2}{6}  \\
\end{array}
\right),
\end{equation}
whose mass eigenvalues are
\begin{eqnarray}
\mathcal{M}_1^2(\phi_0,\psi_0)&=& \frac{1}{2} \left( m_{\phi
}^2+\frac{\lambda _{\phi } \phi_0^2}{2} +\frac{\lambda  \psi_0^2}{2}
+m_{\psi }^2+\frac{\lambda  \phi_0^2}{2}+\frac{\lambda _{\psi }
  \psi_0^2}{2}\right) \nonumber \\ &+&  \frac{1}{2}\sqrt{ \left(
  m_{\phi }^2+\frac{\lambda _{\phi } \phi_0^2}{2} +\frac{\lambda  \psi
    _0^2}{2} -m_{\psi }^2-\frac{\lambda  \phi _0^2}{2}-\frac{\lambda
    _{\psi } \psi_0^2}{2}\right)^2 + 4 \lambda^2 \phi_0^2 \psi_0^2},
\\ \mathcal{M}_2^2(\phi_0,\psi_0)&=& \frac{1}{2} \left( m_{\phi
}^2+\frac{\lambda _{\phi } \phi_0^2}{2} +\frac{\lambda  \psi _0^2}{2}
+m_{\psi }^2+\frac{\lambda  \phi _0^2}{2}+\frac{\lambda _{\psi }
  \psi_0^2}{2}\right) \nonumber \\ &-&  \frac{1}{2}\sqrt{ \left(
  m_{\phi }^2+\frac{\lambda _{\phi } \phi_0^2}{2} +\frac{\lambda  \psi
    _0^2}{2} -m_{\psi }^2-\frac{\lambda  \phi _0^2}{2}-\frac{\lambda
    _{\psi } \psi_0^2}{2}\right)^2 + 4 \lambda^2 \phi_0^2 \psi_0^2},
\\ \mathcal{M}_3^2(\phi_0,\psi_0)&=&  m_{\phi }^2+\frac{\lambda _{\phi
  } \phi_0^2}{6} +\frac{\lambda  \psi_0^2}{2},
\\ \mathcal{M}_4^2(\phi_0,\psi_0)&=&  m_{\psi }^2+\frac{\lambda _{\psi
  } \psi_0^2}{6}+\frac{\lambda  \phi_0^2}{2}.
\end{eqnarray}
\end{widetext} 
When substituting the background fields by their tree-level vacuum
expectation values, 
\begin{eqnarray}
&& \phi_0^2= \frac{-6\lambda_\psi m_\phi^2+18 \lambda
    m_\psi^2}{\lambda_\phi \lambda_\psi-9 \lambda ^2},\\ && \psi_0^2=
  \frac{-6\lambda_\phi m_\psi^2+18 \lambda  m_\phi^2}{\lambda_\phi
    \lambda_\psi-9 \lambda ^2},
\end{eqnarray}
we can recognize that $\mathcal{M}_1$ and $\mathcal{M}_2$ are the two
Higgs modes, associated  with the $\phi_1$ and $\psi_1$ real
components for the two complex scalar fields, while
$\mathcal{M}_3(\phi_0,\psi_0)=\mathcal{M}_4(\phi_0,\psi_0)=0$ are the
corresponding two Goldstone modes, associated with the $\phi_2$ and
$\psi_2$ imaginary components of $\phi$ and $\psi$.  The renormalized
one-loop effective potential is then given by
\begin{eqnarray}
V_{\rm eff, R}&=& \frac{m_{\phi }^{2}}{2}\phi_0^{2}+\frac{m_{\psi
  }^{2}}{2}\psi_0^{2}+\frac{\lambda_\phi}{4!}\phi_0^{4}
+\frac{\lambda_\psi}{4!}\psi_0^{4}+\frac{\lambda}{4}
\phi_0^{2}\psi_0^{2} \nonumber \\ &+&\frac{1}{2}\sum_{i=1}^4
Y\left[\mathcal{M}_i(\phi_0,\psi_0)\right],
\label{Veff1loopR}
\end{eqnarray}
where the function $Y$ is given by one of the previous results
[Eq.~(\ref{YT})] when $T\neq 0,\, B=0$, Eq.~(\ref{YBT}), when $T\neq
0,\, B\neq 0$, or Eq.~(\ref{YBT0}) when  $T= 0,\, B\neq 0$, with the
replacement $\Omega \to \mathcal{M}_i$ in those equations.

In the next section we will study numerically the results obtained
using the OPT-PMS and contrast these results with those obtained 
when using perturbation theory and/or using the one-loop
approximation.

\section{ISB and SNR in the OPT-PMS nonperturbative method}
\label{sec5}

Let us start the exploration of the results for ISB and SNR for the
present model in the OPT-PMS nonperturbative method. {}For comparison
purposes and for easiness of analysis, we will divide our presentation
for the cases of $T\neq 0$ and $B=0$, $T=0$ and $B\neq 0$ and, finally
include the combined effects of temperature and external magnetic
field.

\subsection{ISB and SNR in the OPT-PMS nonperturbative method at
  $T\neq 0$ and $B=0$}
\label{sec5T}

\begin{center}
\begin{figure}[!htb]
\subfigure[The background scalar field $\tilde
  \phi$.]{\includegraphics[width=7.5cm]{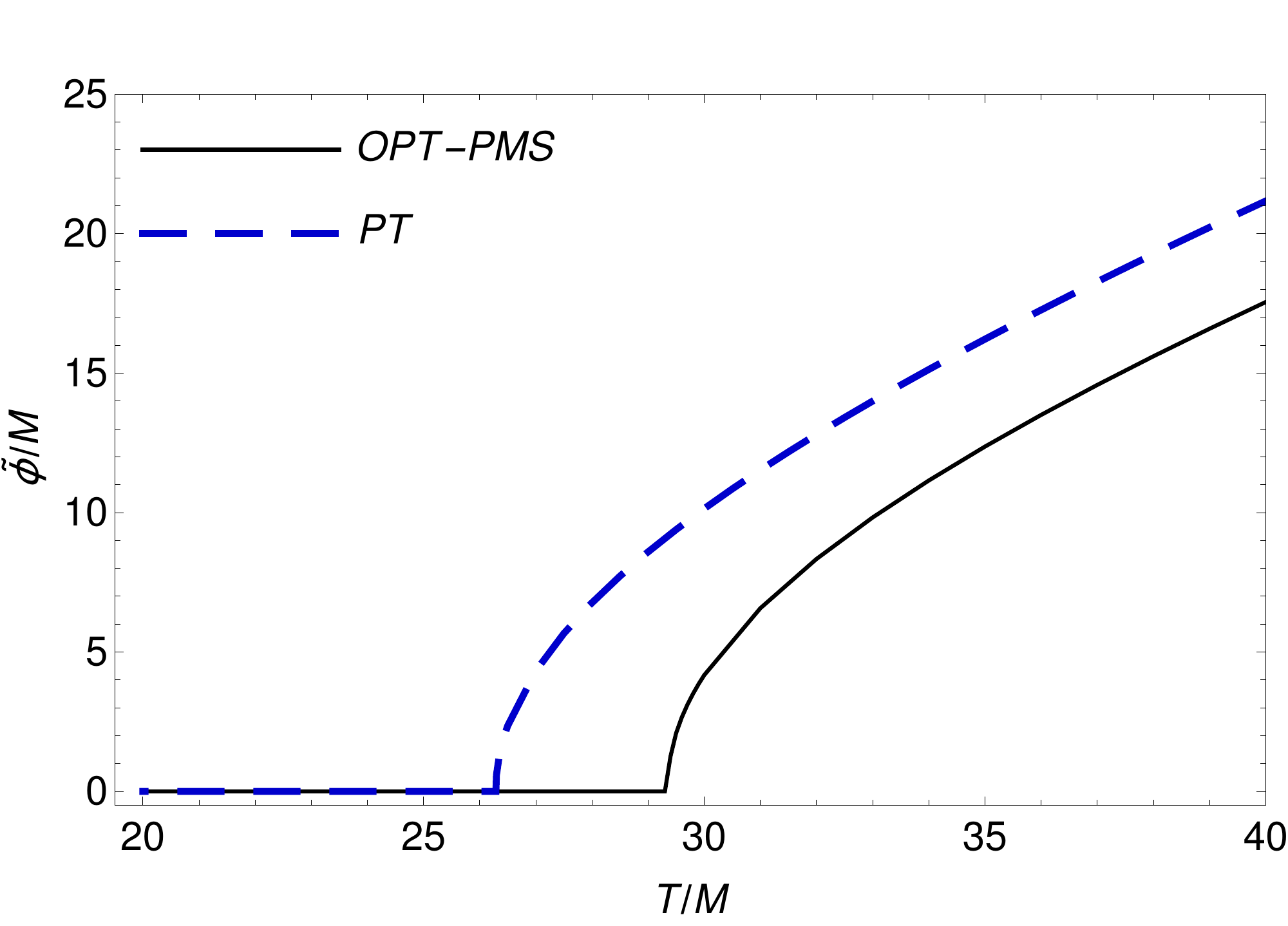}}
\subfigure[The critical temperature for
  ISB.]{\includegraphics[width=7.5cm]{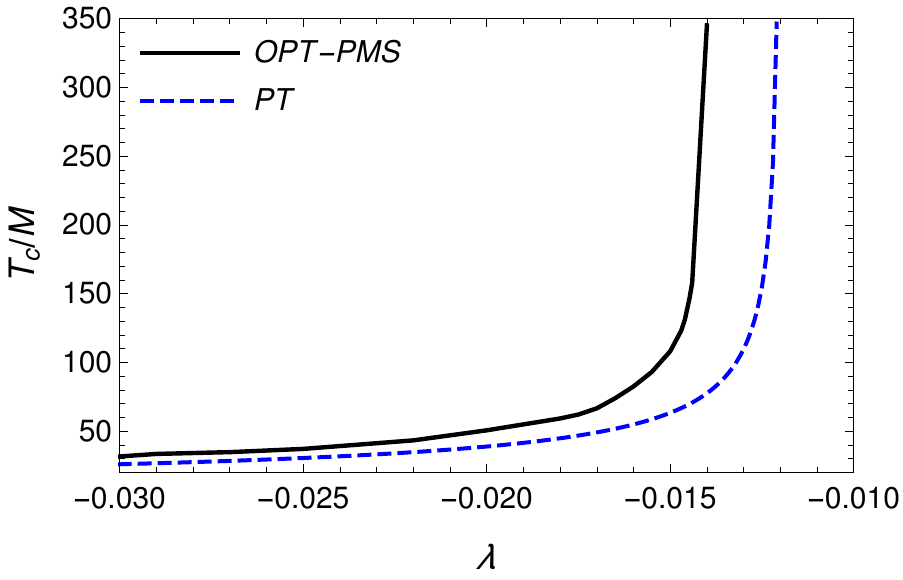}}
\caption{The field expectation value  $\tilde \phi$ as  a function of
  the temperature, panel (a), and the critical temperature for ISB as a function
  of the intercoupling $\lambda$, panel (b).  Both OPT and PT are considered for
  comparison.  The parameters considered are: $m_\phi^2 = m_\psi^2 =
  M^2 >0$, $\lambda_\phi=0.018$, $\lambda_\psi=0.6$ and,  for panel
  (a), $\lambda=-0.03$.}
\label{fig1}
\end{figure}
\end{center}

In the case of only including finite-temperature effects, we need
the thermal functions defined in Sec.~\ref{casea}. The respective 
momentum integrals are solved numerically directly. We start by
considering the case with $m_\phi^2 > 0$ and $m_\psi^2> 0$. {}For
illustration purposes, we choose the couplings such that it predicts
ISB in the $\phi$ direction, while in the $\psi$ direction the system
remains in the symmetry restored phase. 

In the example considered in
{}Fig.~\ref{fig1}(a) there is an ISB transition in the
$\phi$ direction at a critical temperature $T_{c,\phi}/M \simeq 29.3$
in the OPT-PMS case, while for PT the critical temperature is
smaller, $T_{c,\phi}/M \simeq 26.1$. In
{}Fig.\ref{fig1}(b) it is shown how the critical temperature for ISB
in each case behaves as a function of the intercoupling $\lambda$.
Note that ISB tends to disappear at less negative values of
$\lambda$ (which is a consequence of approaching the boundedness condition
for the couplings).  This is signaled by a
diverging behavior of $T_{c}$, which happens at a smaller (in modulus) value 
for $\lambda$ in the PT case as compared to the OPT.

\begin{center}
\begin{figure}[!htb]
\subfigure[The effective masses Eqs.~(\ref{mphieff}) and
  (\ref{mpsieff})]{\includegraphics[width=7.5cm]{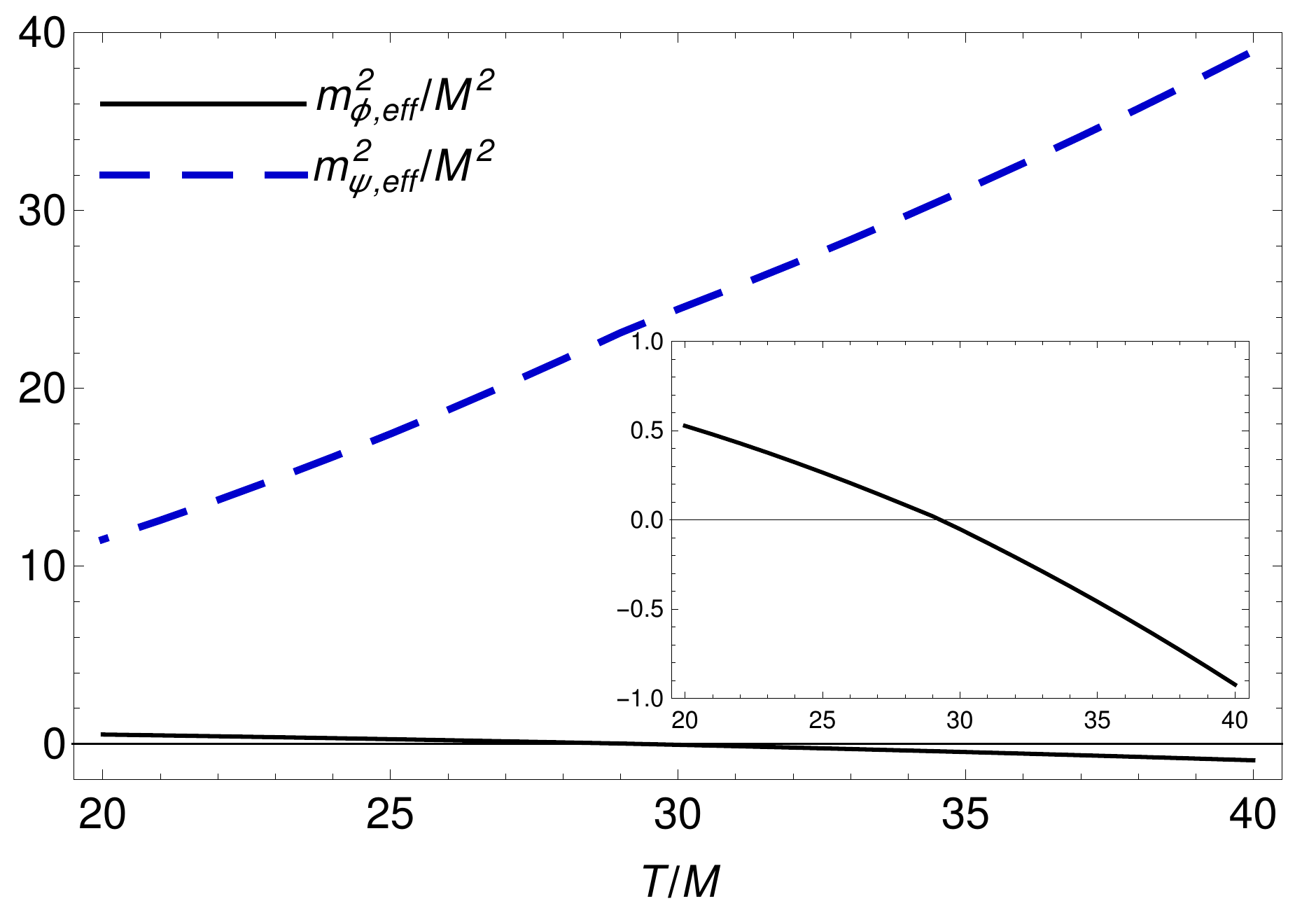}}
\subfigure[The physical masses for the fields, Eqs.~(\ref{mphi1eff}),
  (\ref{mphi2eff}), (\ref{mpsi1eff}) and
  (\ref{mpsi2eff}).]{\includegraphics[width=7.5cm]{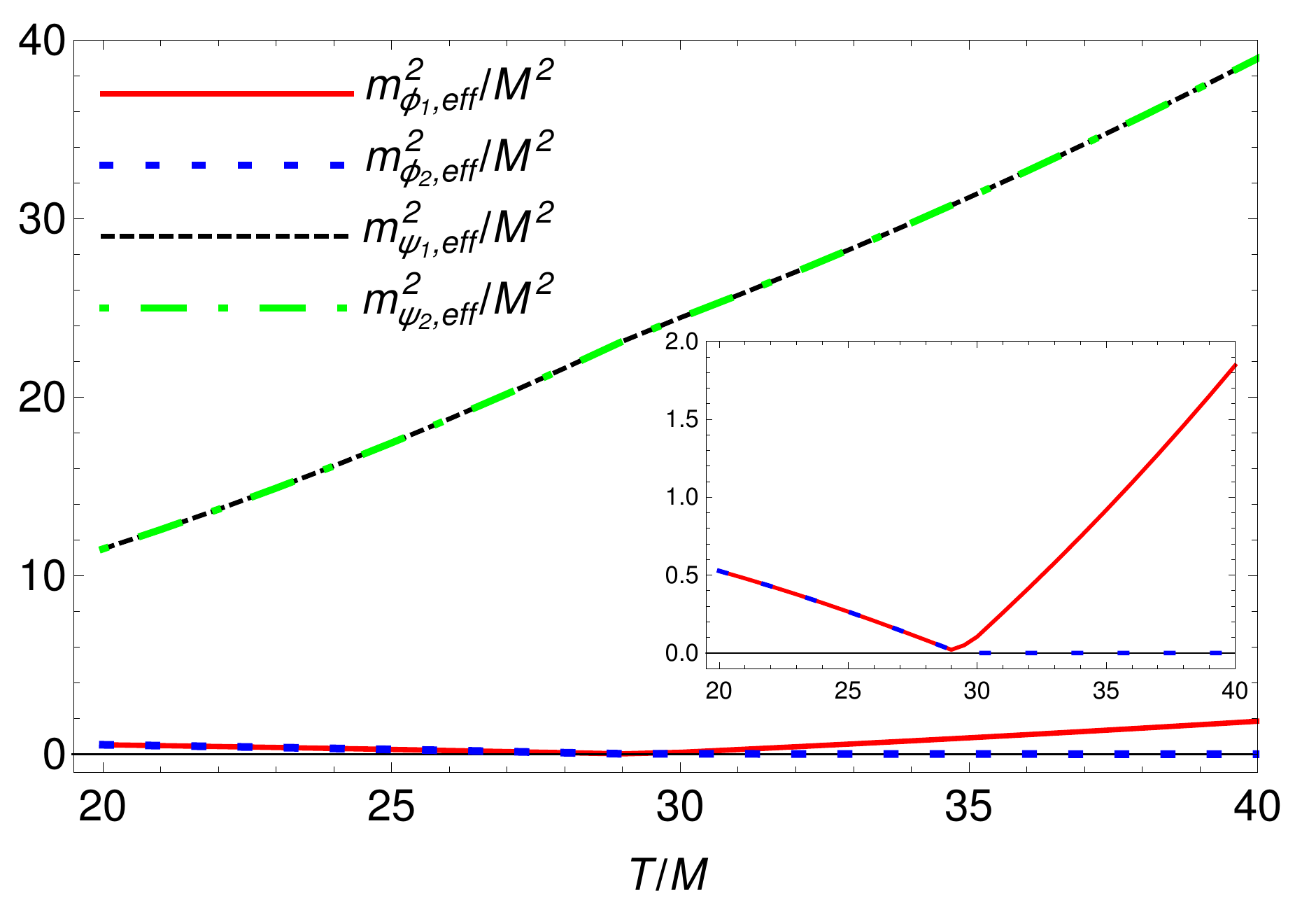}}
\caption{The curvature of the effective potential in the direction of
  the $\phi$ and $\psi$ fields, panel (a), the physical masses
  corresponding to the Higgs and Goldstone modes for each of the
  fields, panel (b). The parameters considered are: $m_\phi^2 =
  m_\psi^2 = -M^2 >0$, $\lambda_\phi=0.018$, $\lambda_\psi=0.6$, and
  $\lambda=-0.03$. The insets in the plots help to better see the behavior
for the effective mass in the direction of $\phi$ [panel (a)]
and for the Higgs and Goldstone modes for the $\phi$ field [panel (b)].}
\label{fig2}
\end{figure}
\end{center}

We consider the same conditions used in {}Fig.\ref{fig1} 
where we have seen ISB in the $\phi$ direction, while $\psi$ remains in its symmetry restored
phase, also in {}Fig.~\ref{fig2}.  In {}Fig.~\ref{fig2} we now show the effective 
masses Eqs.~(\ref{mphieff}) and (\ref{mpsieff}) and we contrast them with the 
physical ones, given by Eqs.~(\ref{mphi1eff}),
(\ref{mphi2eff}), (\ref{mpsi1eff}), and (\ref{mpsi2eff}) in the OPT-PMS scheme. We 
can see from {}Fig.~\ref{fig2}(a) that at the transition point the effective mass 
square $m^{2}_{\phi, \rm{eff}}$ changes from positive (symmetry restored) to negative 
(symmetry breaking), while $m^{2}_{\psi, \rm{eff}}$ remains positive throughout 
the temperature range shown, indicating that the symmetry remains restored in 
that direction. In {}Fig.\ref{fig2}(b) we see that the Goldstone theorem applies, 
as it should, for each of the field directions. In this case, both Higgs and 
Goldstone modes agree with each other for all range of temperature in the $\psi$ 
direction. {}For the $\phi$ direction, the Higgs- and the Goldstone-like masses 
agree with each other in  the $\phi$ direction for temperatures  $T\leq T_{c,\phi}$.

\begin{center}
\begin{figure}[!htb]
\subfigure[The background fields $\tilde{\phi}$ and $\tilde{\psi}$.]
{\includegraphics[width=7.5cm]{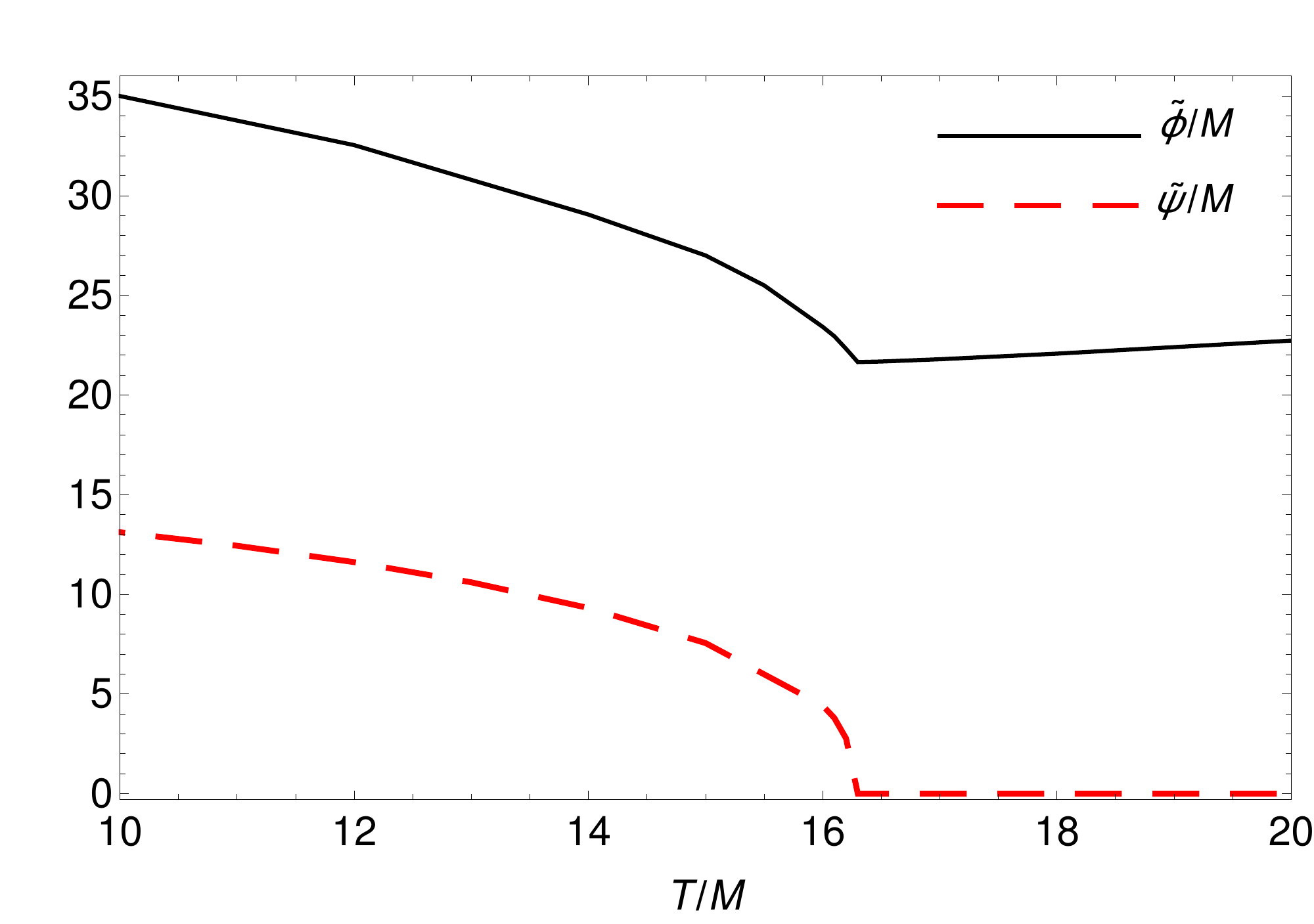}}
\subfigure[The effective masses Eqs.~(\ref{mphieff}) and
  (\ref{mpsieff}).]{\includegraphics[width=7.5cm]{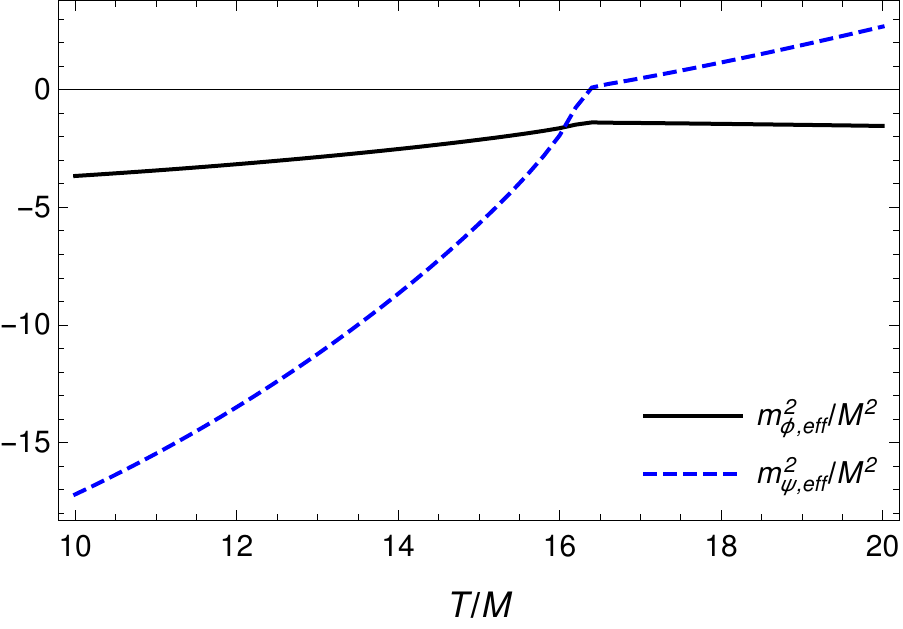}}
  \subfigure[The physical masses for the fields, Eqs.~(\ref{mphi1eff}),
  (\ref{mphi2eff}), (\ref{mpsi1eff}) and
  (\ref{mpsi2eff}).]{\includegraphics[width=7.5cm]{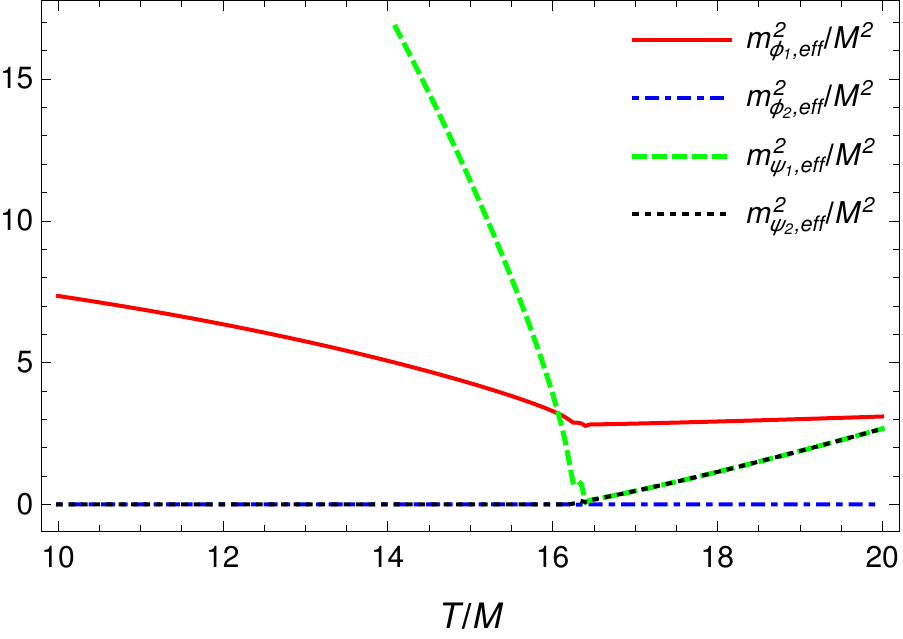}}
\caption{The background fields $\tilde{\phi}$ and $\tilde{\psi}$ as a 
function of temperature, panel (a), 
the curvature of the effective potential in the direction of
  the $\phi$ and $\psi$ fields, panel (b), and the physical masses
  corresponding to the Higgs and Goldstone modes for each of the
  fields, panel (c).  The parameters considered in these figure are $m_\phi^2 =
  m_\psi^2 = -M^2 <0$, $\lambda_\phi=0.018$, $\lambda_\psi=0.6$, and
  $\lambda=-0.03$.}
\label{fig3}
\end{figure}
\end{center}

Now, let us consider the case of SNR. Hence, we consider  $m_\phi^2 <
0$ and $m_\psi^2 < 0$, i.e., we start with the system in the symmetry
broken state in both the field directions. In {}Fig.~\ref{fig3}(a) we
show both background fields $\tilde \phi$ and $\tilde \psi$ as a
function of the temperature in the OPT case. The parameters chosen are
such that  $m_\phi^2 = m_\psi^2 = -M^2 < 0$, with the same couplings
as in {}Fig.~\ref{fig1}, $\lambda_\phi=0.018$, $\lambda_\psi=0.6$, and
$\lambda=-0.03$. {}For these parameters the symmetry remains broken,
i.e., there is SNR in the direction of $\phi$,  while there is a
symmetry restoration at a critical temperature in the direction of
$\psi$.   {}For the parameters considered in this example, we have that
$T_{c,\psi}/M \simeq 16.4$ in the OPT case, while in the one-loop
approximation we find $T_{c,\psi}/M \simeq 15.9$. 
In this same example showing SNR, 
it is useful to give the effective
masses Eqs.~(\ref{mphieff}) and (\ref{mpsieff}) and to contrast them
with the physical ones, given by Eqs.~(\ref{mphi1eff}),
(\ref{mphi2eff}), (\ref{mpsi1eff}) and (\ref{mpsi2eff}) in the OPT
scheme. These are shown in {}Figs.~\ref{fig3}~(b) and 3(c), respectively.
We can clearly see from
{}Fig.~\ref{fig3}(b) that at the transition point the effective mass
square $m_{\psi,{\rm eff}}^2$ changes from negative (symmetry breaking)
to positive (symmetry restored) at the critical temperature
$T_{c,\psi}/M \simeq 16.4$, while $m_{\phi,{\rm eff}}^2$ remains
negative throughout the temperature range shown, indicating SNR in the
direction of $\phi$. In {}Fig.~\ref{fig3}(c) we see that the Goldstone
theorem applies, as it should, for each of the field directions. The
physical masses remain positive definite as also expected.  In
particular, after symmetry restoration (in the direction of $\psi$),
both the Higgs and Goldstone modes agree with each other,
$m_{\psi_1,{\rm eff}} = m_{\psi_2,{\rm eff}}$ for  $T\geq T_{c,\psi}$.

\subsection{ISB and SNR in the OPT-PMS nonperturbative method at
  $T= 0$ and $B\neq 0$}
\label{sec5B}

\begin{center}
\begin{figure}[!htb]
\subfigure[The background scalar field $\tilde
  \psi$.]{\includegraphics[width=7.5cm]{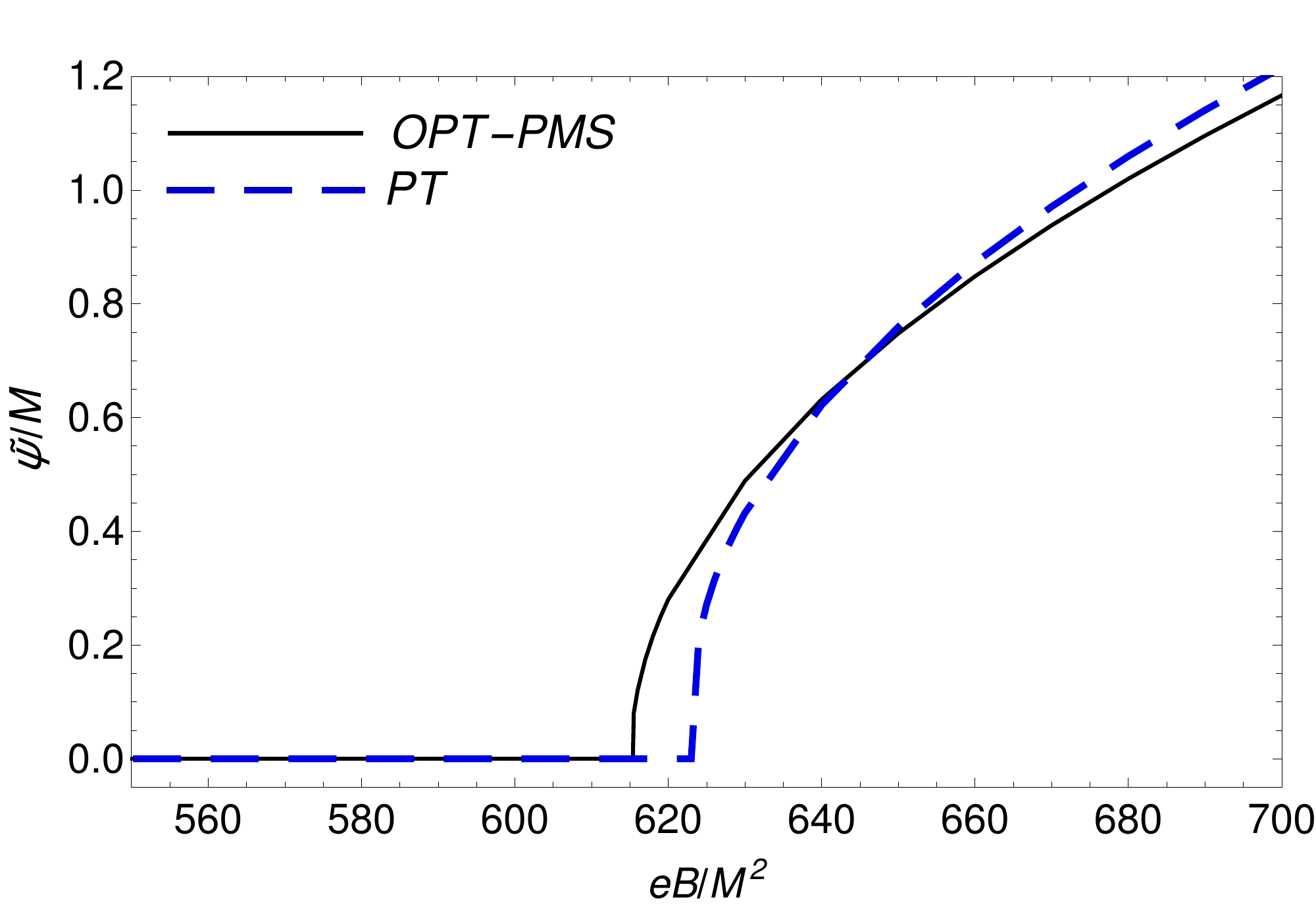}}
\subfigure[The critical magnetic field for
  ISB.]{\includegraphics[width=7.5cm]{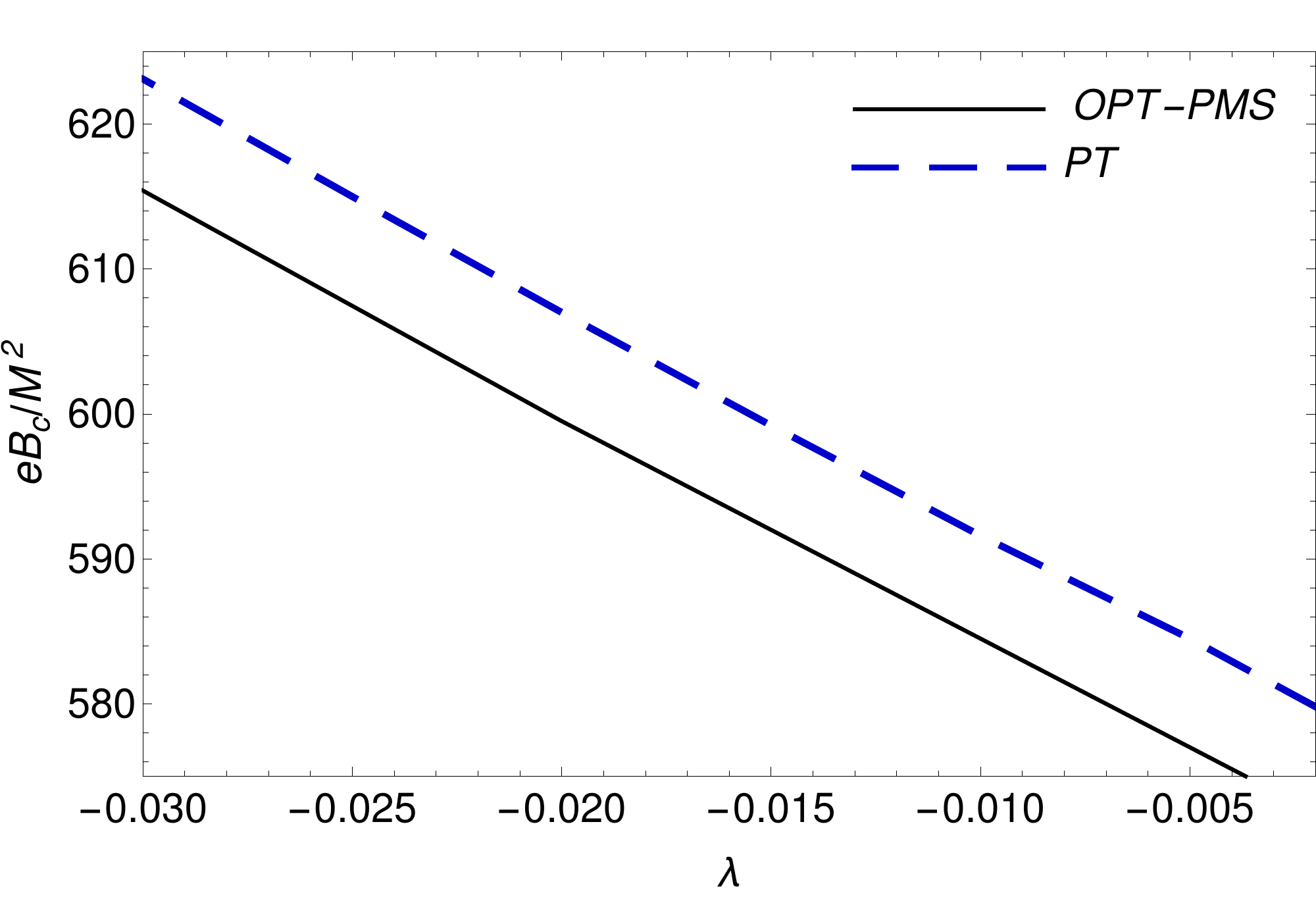}}
\caption{The field expectation value  $\tilde \psi$ as  a function of
  the magnetic field and the critical magnetic field for ISB as a
  function of the intercoupling $\lambda$. Both OPT and PT are
  considered for comparison.  The parameters considered are $m_\phi^2
  = m_\psi^2 = M^2 >0$, $\lambda_\phi=0.018$, $\lambda_\psi=0.6$ and,
  for panel (a), $\lambda=-0.03$.}
\label{fig4}
\end{figure}
\end{center}

Turning now to the case of zero temperature, but finite external
magnetic field, we start by looking at the results in the case when
$m_\phi^2> 0$ and $m_\psi^2 >0$, i.e.,  starting in the symmetry
restored phase for both $\phi$ and $\psi$. The couplings are again
chosen such that $2 \lambda_\phi/3 + \lambda <0$ (with $\lambda
<0$). A representative example of this case is shown in
{}Fig.~\ref{fig4}. As expected, the situation gets reversed with
respect to what has been shown in the corresponding case at finite
temperature but zero external magnetic field and shown in
{}Fig.~\ref{fig1}.  Here, we have ISB in the direction of the $\psi$
field, while the symmetry remains restored in the $\phi$ direction,
i.e., $\tilde \phi=0$ throughout the range of magnetic field
values considered.  In the direction of the $\psi$ field, the critical
magnetic field, for the case of the parameters
$\lambda_\phi=0.018$, $\lambda_\psi=0.6$, and $\lambda=-0.03$ is given
by  $eB_{c,\psi}/M^2 = 615.1$ in the case of the OPT, while in PT we
obtain that  $eB_{c,\psi}/M^2 = 623.1$. 

\begin{center}
\begin{figure}[!htb]
\subfigure[The background scalar field $\tilde
  \phi$.]{\includegraphics[width=7.5cm]{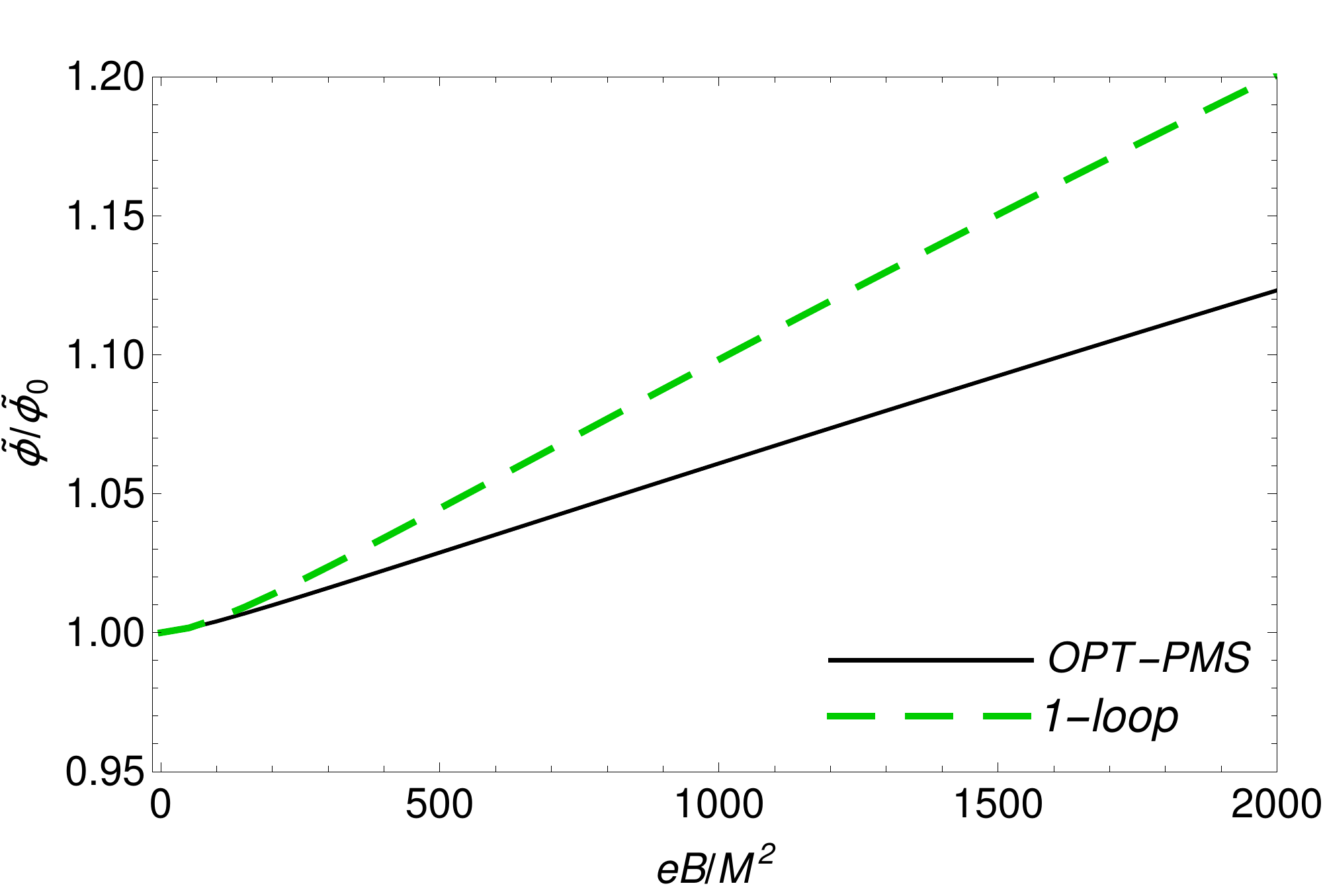}}
\subfigure[The background scalar field $\tilde
  \psi$.]{\includegraphics[width=7.5cm]{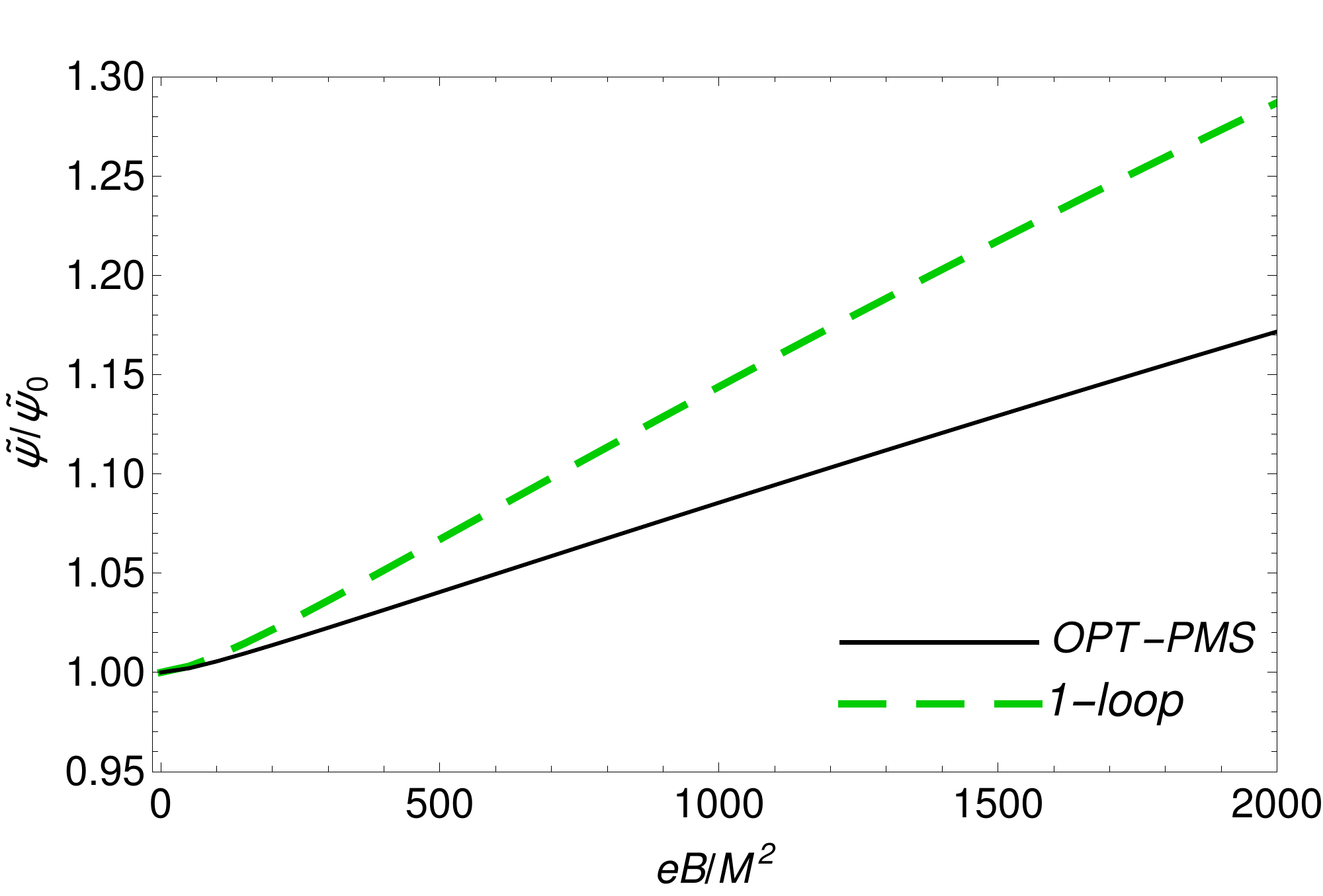}}
\caption{The field expectation values  $\tilde \phi$ and $\tilde \psi$
  as  a function of the magnetic field. Here,  both OPT and the
  one-loop approximation are considered for comparison.  The
  parameters considered are: $m_\phi^2 = m_\psi^2 = -M^2 <0$,
  $\lambda_\phi=0.018$, $\lambda_\psi=0.6$, and  $\lambda=-0.03$.
{}For convenience, the fields were normalized by their
respective vacuum expectation values at $B=0$.}
\label{fig5}
\end{figure}
\end{center}

Let us now consider the case of starting in the symmetry broken phase
for both field directions, i.e.,  taking $m_\phi^2< 0$ and $m_\psi^2
<0$. Again we assume that $2 \lambda_\phi/3 + \lambda <0$.  The
results in this case are shown in  {}Fig.~\ref{fig5}. Note here that
both $\phi$ and $\psi$ remain  in a symmetry broken state as a
consequence of the magnetic field favoring symmetry breaking. This is
akin to the magnetic catalysis effect seen in general due to a
magnetic field, which tends to  enhance the symmetry
breaking~\cite{Shovkovy:2012zn}. In  {}Fig.~\ref{fig5} we have
considered both OPT and the one-loop approximation for comparison
purposes.

\subsection{ISB and SNR in the OPT-PMS nonperturbative method at
  $T \neq 0$ and $B\neq 0$}
\label{sec5C}

Let us now consider the effects from both temperature and external
magnetic field. As already observed, this is a particularly
interesting case, since there is a competition between  the thermal
effects and the magnetic field, which one acting in an opposite direction
as far as symmetry breaking and restoration are concerned.
Before entering in our numerical results for this case, let us recall
that in the expressions involving the magnetic field, e.g.,
Eqs.~(\ref{JBTB}) and (\ref{IBTB}), the thermal integrals also require
a sum over the Landau levels. While at large magnetic field values,
$eB \gg \Omega^2$, one only requires to sum over a few Landau levels for
proper convergence of the expressions, in the weak magnetic field
regime,  $eB \ll \Omega^2$, one typically requires to consider a large
number of levels in the sum. There have been a few, but very reliable approaches
in the literature to handle this issue in the weak magnetic field
regime.  {}For example, Ref.~\cite{Duarte:2011ph} made use
of the Euler-Maclaurin formula as a way to work out the sum over the
large number of Landau's levels. In Ref.~\cite{Ayala:2004dx}, the
authors have proposed instead a weak magnetic field approximation for
the bosonic propagator.  In this case, the propagator in Euclidean
spacetime for the charged scalar fields in the presence of an external
magnetic field is approximated as~\cite{Ayala:2004dx} 
\begin{eqnarray}
&& \frac{1}{\omega_{n}^{2}+E^{2}({\bf p})} \to \nonumber \\ &&
  \frac{1}{\omega_{n}^{2}+E^{2}({\bf p})} \left\{ 1 - \frac{(eB)^{2}}{
    \left[\omega_{n}^{2}+E^{2}({\bf p})\right]^{2} }  +\frac{
    2\ (eB)^{2}\ p_{\bot}^{2} }{ \left[ \omega_{n}^{2}+E^{2}({\bf
        p})\right]^{3} }   \right\}, \nonumber \\
\label{weakBprop}
\end{eqnarray}
where $E^{2}({\bf p}) ={\bf p}^2+ \Omega^{2}$, with ${\bf
  p}^2=p_{z}^{2}+ p_{\bot}^{2}$ and  $p_{\bot}^2 = p_{x}^{2}+p_{y}^{2}$.
In this work, we have made use of both the  Euler-Maclaurin formula
considered in Ref.~\cite{Duarte:2011ph}  as also the weak magnetic
field approximation for the bosonic propagator given by
Eq.~(\ref{weakBprop}).  In the weak magnetic field regime $eB \ll
\Omega^2$ both approaches are found to agree quite well.  {}For the
case of strong magnetic fields, we have considered a sufficient number of
Landau's levels such to have convergence for the results. 
With these due cares taken into account, we now present our results.

\begin{center}
\begin{figure}[!htb]
\subfigure[The background scalar field $\tilde
  \phi$.]{\includegraphics[width=7.5cm]{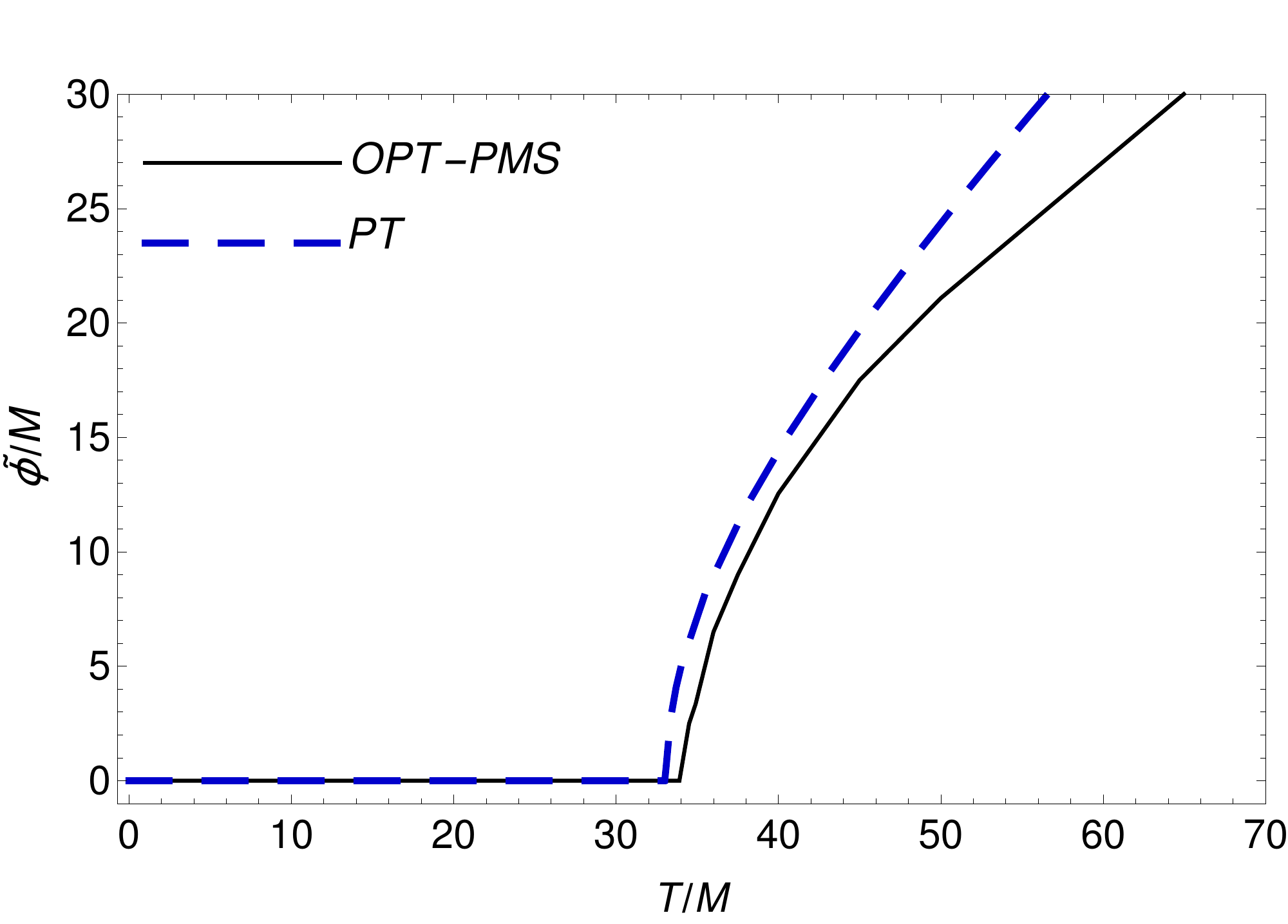}}
\subfigure[The critical temperature for ISB as a function of the
  magnetic field.]{\includegraphics[width=7.5cm]{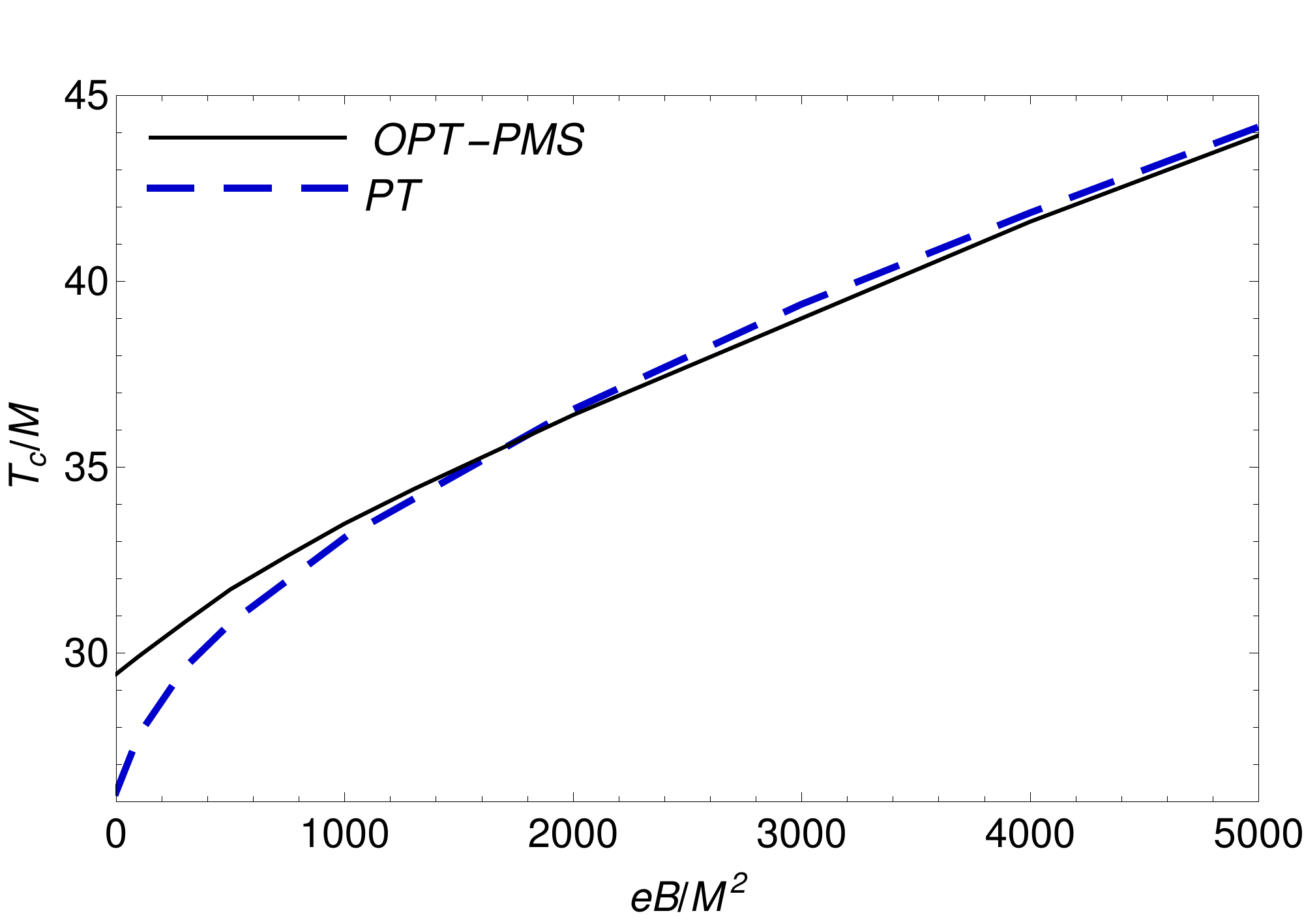}}
\caption{The field expectation value  $\tilde \phi$ as  a function of
  the temperature at a fixed external magnetic field $eB/M^2 = 10^3$,
  panel (a), and the critical temperature for ISB in the
  $\phi$ direction as a function of magnetic field, panel (b). Both
  OPT and PT are considered for comparison.  The parameters considered
  are: $m_\phi^2 = m_\psi^2 = M^2 >0$, $\lambda_\phi=0.018$,
  $\lambda_\psi=0.6$, and $\lambda=-0.03$.}
\label{fig6}
\end{figure}
\end{center}

In {}Fig.~\ref{fig6} we explore the effect of the external
magnetic field in the ISB case for the $\phi$ field. Note that from
the result shown in {}Fig.~\ref{fig6}(a), PT exhibits a much stronger
departure from the OPT result for the background field $\tilde \phi$
as the temperature increases beyond the critical temperature for ISB
when in a strong magnetic field regime. In {}Fig.~\ref{fig6}(b) we
show how the critical temperature  for ISB in this example changes
with an increasing magnetic field. One notices that for very strong
fields, the critical point is obtained in the OPT and PT approximations as
they approach each other, showing that nonperturbative effects brought by
the OPT tend to be less important.

\begin{center}
\begin{figure}[!htb]
\subfigure[The background scalar field $\tilde
  \phi$.]{\includegraphics[width=7.5cm]{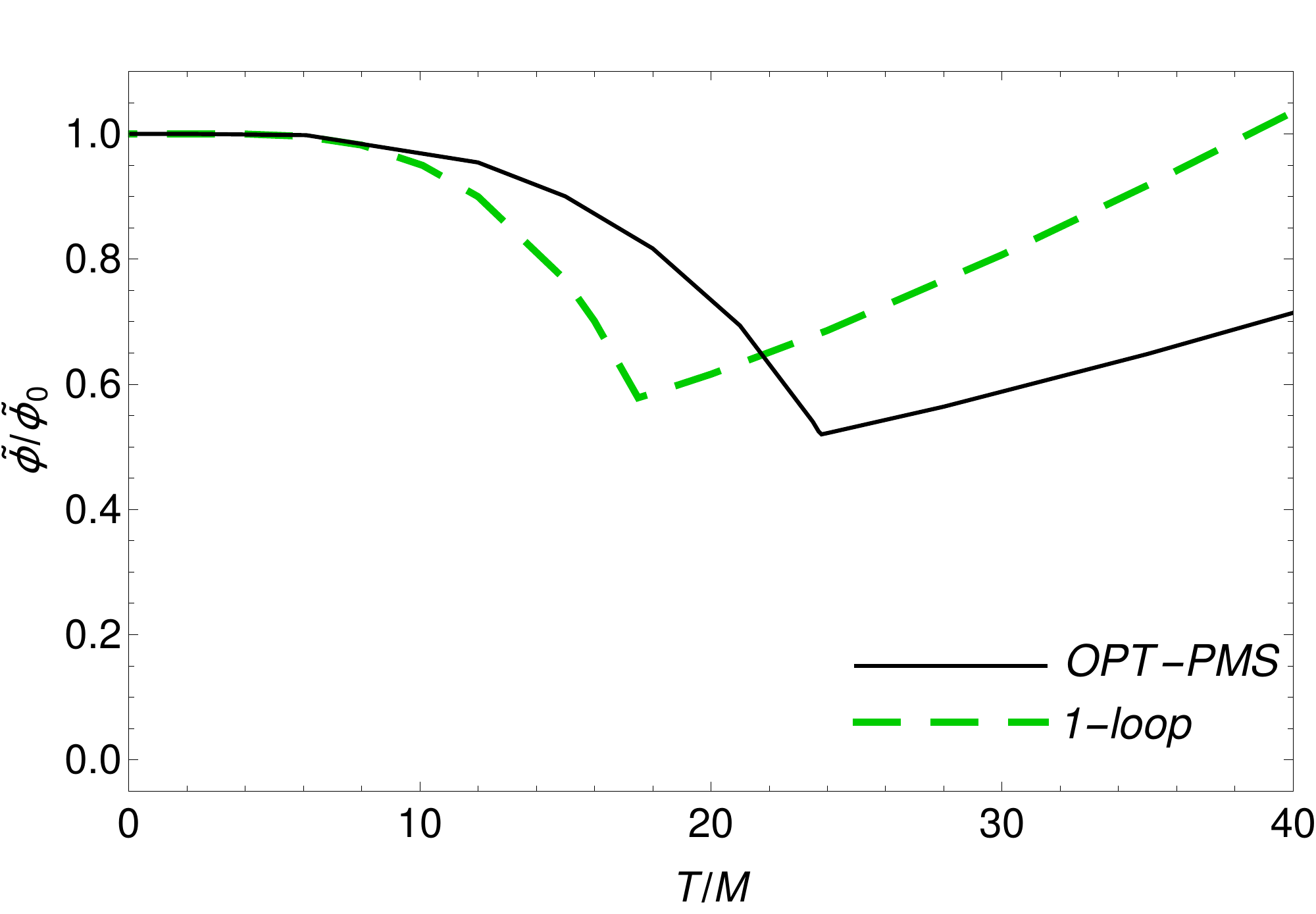}}
\subfigure[The background scalar field $\tilde
  \psi$.]{\includegraphics[width=7.5cm]{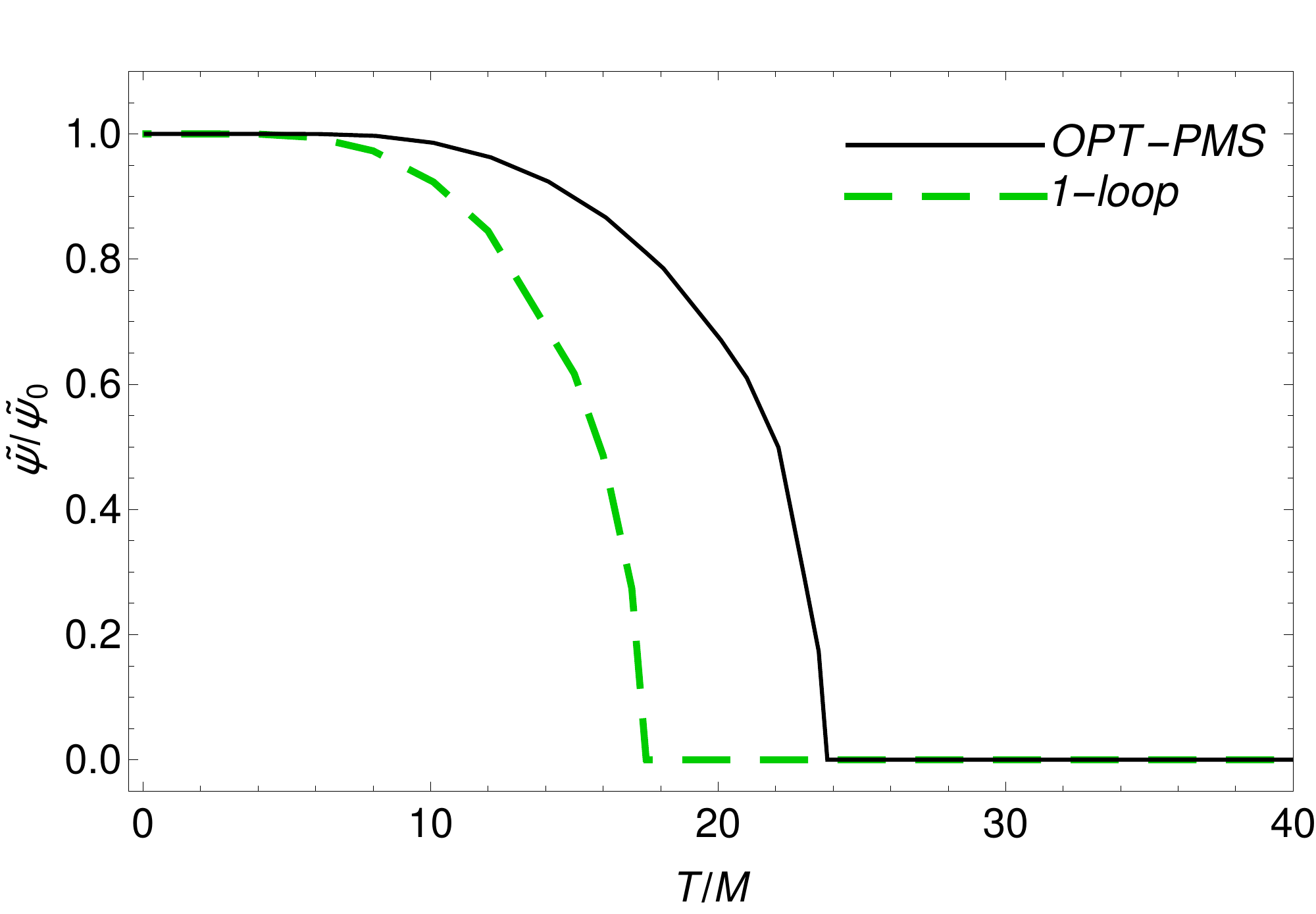}}
\caption{The field expectation values  $\tilde \phi$, panel (a), and
  $\tilde \psi$, panel (b),  as  a function of the temperature at a
  fixed value of the magnetic field, $eB/M^2 = 10^3$.  Here,  both OPT
  and the one-loop approximation are considered for comparison.  The
  parameters considered are $m_\phi^2 = m_\psi^2 = -M^2 <0$,
  $\lambda_\phi=0.018$, $\lambda_\psi=0.6$, and  $\lambda=-0.03$.
{}For convenience, the fields were normalized by their
respective values at $T=0$.}
\label{fig7}
\end{figure}
\end{center}

\begin{center}
\begin{figure}[!htb]
\includegraphics[width=7.5cm]{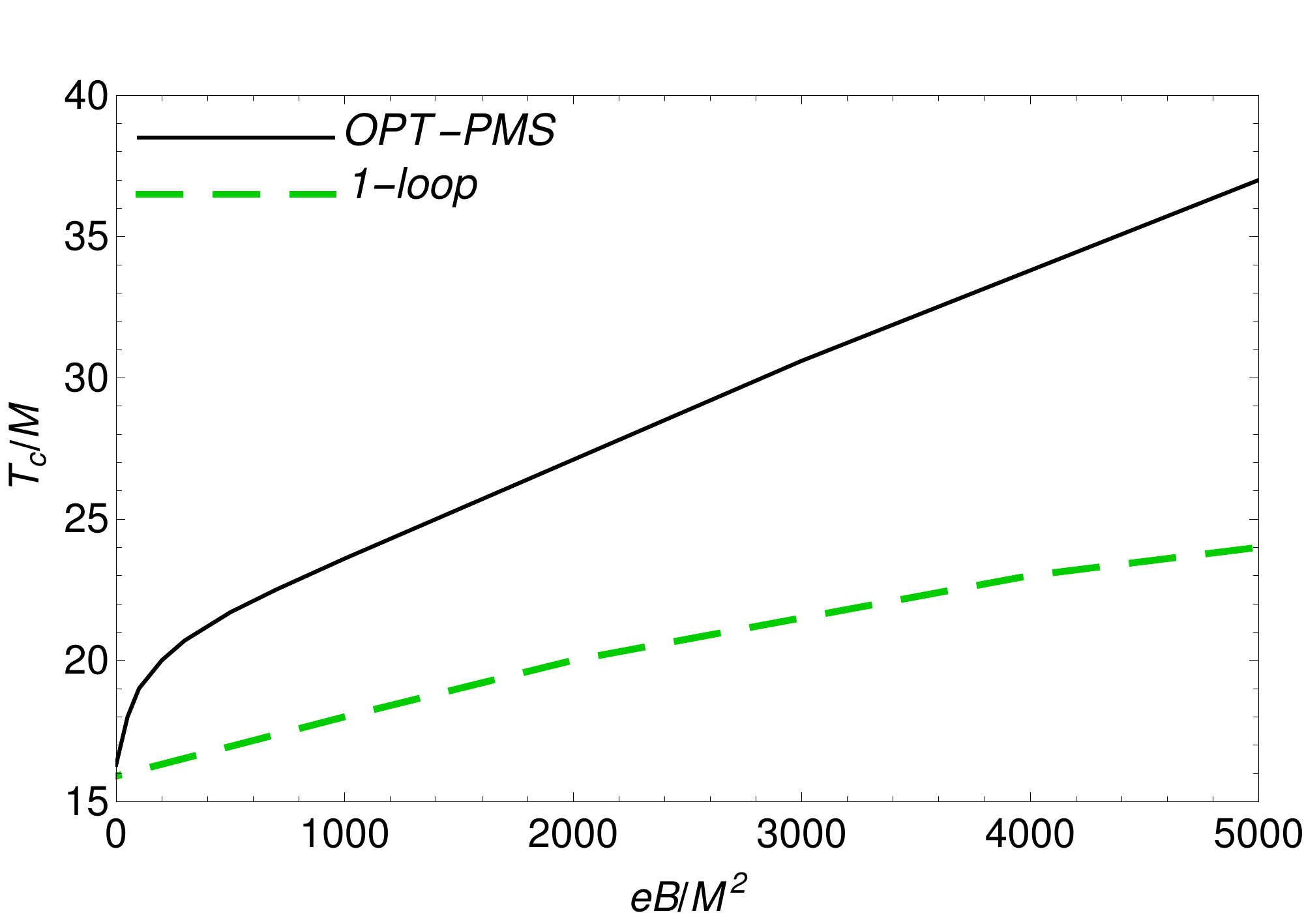}
\caption{The critical temperature for symmetry restoration in the
  $\tilde \psi$-field direction as a function of the magnetic field.
  Here,  both OPT and the one-loop approximation are considered for
  comparison.  The parameters considered are $m_\phi^2 = m_\psi^2 =
  -M^2 <0$, $\lambda_\phi=0.018$, $\lambda_\psi=0.6$, and
  $\lambda=-0.03$.}
\label{fig8}
\end{figure}
\end{center}

{}Finally, in {}Fig.~\ref{fig7}, we study the SNR case when in the
presence of strong magnetic fields.  Here, we compare again the
results from the OPT with the one-loop approximation and the parameters
are chosen such that there is symmetry restoration in the direction of
the $\psi$ field, while the symmetry remains broken (SNR) in the
direction of $\phi$. In the $\psi$ direction and for an external
magnetic field of $eB/M^2=10^3$, the critical temperature for symmetry
restoration in the direction of $\psi$ is found to be $T_{c, \psi}/M
\simeq 23.9M$ in the OPT case, while in the one-loop  approximation we
find that $T_{c,\psi}/M \simeq 17.7$. The way the critical temperature
$T_{c,\psi}$ changes in both cases as a function of the external
magnetic field is shown in  {}Fig.~\ref{fig8}.  Here, opposite to
the previous case of ISB shown in {}Fig.~\ref{fig6}(b), the one-loop
approximation tends to underestimate the critical temperature when
compared to the OPT result. The difference between the results also
increases the larger is the magnetic field.

\section{Conclusions}
\label{conclusions}

In this work, we have shown that the introduction of an external
magnetic field can induce significant changes in the phase  structure
of a system composed by two complex scalar fields with both 
self- and inter-interactions.  This is a consequence of the fact that thermal effects
tend to act oppositely to those obtained for an external magnetic field. This
situation gets particularly more involved whenever there is a
possibility of ISB and/or SNR phenomena, which results by an appropriate
choice for the parameters of the model, e.g., the coupling constants,
in particular for the intercoupling between the fields. 

In our study we have used the effective potential for the analysis
of the phase structure of the model.  The derivation of the effective
potential and the results obtained from it were performed considering
the nonperturbative method of the OPT. The results obtained in the OPT method were compared to
those obtained in the perturbation theory as well as with those
obtained when using the one-loop
approximation for the effective potential.

Our results show that the effect of the magnetic field tends to always
increase the critical temperature, making symmetry breaking more
easily achieved whenever symmetry restoration is involved.  In the
case of ISB, the critical temperature also increases with the magnetic
field, although here we can  interpret that the external magnetic field
makes it more difficult to achieve ISB, thus producing a higher $T_c$.

In summary, our results still corroborate with the existence of
ISB/SNR, even when including the effects of an external magnetic field
and in the context of the nonperturbative OPT scheme.  In the absence of
a magnetic field, the critical
temperature in the OPT case tends always to be larger than in the
perturbation theory, or in the one-loop approximation, for both ISB and SNR. 
However, in the absence of thermal effects, but in a finite magnetic field,
the critical magnetic field tends to be smaller in the OPT than in the
perturbative and loop approximation cases.
When both the external magnetic field and thermal effects are present, the differences 
between the OPT and the perturbative and loop approximation cases
vary according to the magnitude of the external magnetic field.
Nevertheless, the presence of an external magnetic field, due to the magnetic catalysis 
effect, pushes the energy scale
for phase transition at finite temperature in both cases to be larger than in the absence of
external fields. 

As a possible future extension of this work, one could for example try to study
the issue of inverse magnetic catalysis in the decoupled case, i.e., when $\lambda=0$,
in the OPT.  Inverse magnetic catalysis seems only to be possible to be
realized in a nonperturbative context~\cite{Bandyopadhyay:2020zte}. 
In particular, by having effective
coupling constants that include both thermal and magnetic field effects, 
it has been shown to be able to manifest inverse magnetic catalysis, as, e.g., in the model
of a complex scalar field studied in Ref.~\cite{Ayala:2014iba}.
A similar setting could possibly also be studied in the context of the OPT, 
but likely only when carrying out the calculations up to at least 
second order in the OPT, such that vertices (couplings) could also 
receive thermal and magnetic field effects. This is an interesting but
much more technical and difficult problem that we hope to address in the
future.

\appendix

\section{Renormalization}
\label{appA}

In the OPT scheme one needs only the standard renormalization terms
for fully renormalizing the effective potential for the model, e.g.,
vacuum, masses, and coupling constant counterterms. These counterterms 
need then to be
derived at the appropriate order in the OPT method.  In the present
case, we have derived the effective potential up to order $\delta$ in
the OPT scheme, which implies requiring only a vacuum and appropriate
masses counterterms [counterterms for the couplings are only necessary
when carrying out the derivation at ${\cal O}(\delta^2)$ and higher].

To fully renormalize the model at  ${\cal O}(\delta)$  we need the
mass counterterms:

\begin{eqnarray}
\Delta m_\phi &=&
\frac{2\delta\lambda_{\phi}}{3}\frac{\Omega_{\phi}^2}{(4\pi)^2}
\frac{1}{\epsilon}+
\delta \lambda\frac{\Omega_{\psi}^2}{(4\pi)^2}\frac{1}{\epsilon},
\\ \Delta m_\psi &=&\frac{2\delta\lambda_{\psi}}{3}
\frac{\Omega_{\psi}^2}{(4\pi)^2}\frac{1}{\epsilon}+
\delta\lambda\frac{\Omega_{\phi}^2}{(4\pi)^2}\frac{1}{\epsilon}.
\label{CTMassa}
\end{eqnarray} 
These two mass counterterms, in particular, generate the last two
terms shown in Eq.~(\ref{fullVeff}) that also contribute at
$\mathcal{O}(\delta^{1})$. In particular, note that these
terms are fundamental to remove the crosses divergence with
temperature dependence that appears in the computation of the
two-loop contributions in Eq.~(\ref{fullVeff}).

{}Finally, at the order $\delta$ in the OPT, the renormalization is
completed with the inclusion of the vacuum counterterms,
\begin{eqnarray}
\Delta_{\rm{vac}}&=&\frac{\Omega_{\phi}^4}{2(4\pi)^2}\frac{1}{\epsilon}+
\frac{\Omega_{\psi}^4}{2(4\pi)^2}\frac{1}{\epsilon}  - \delta
\eta_{\phi}^2\frac{\Omega_{\phi}^2}{(4\pi)^2}\frac{1}{\epsilon}-
\delta
\eta_{\psi}^2\frac{\Omega_{\psi}^2}{(4\pi)^2}\frac{1}{\epsilon}\nonumber
\\ &+&
\frac{1}{3}\delta\lambda_{\phi}\frac{\Omega_{\phi}^4}{(4\pi)^4}
\frac{1}{\epsilon^2}
+
\frac{1}{3}\delta\lambda_{\psi}\frac{\Omega_{\psi}^4}{(4\pi)^4}
\frac{1}{\epsilon^2}
+
\delta\lambda\frac{\Omega_{\phi}^2\Omega_{\psi}^2}{(4\pi)^4}
\frac{1}{\epsilon^2}.
\nonumber \\
\label{VacuumCT}
\end{eqnarray}

\section*{ACKNOWLEDGMENTS}

R.L.S.F.~ is partially supported by Conselho Nacional de
Desenvolvimento Cient\'ifico e Tecnol\'ogico  (CNPq), Grant
No. 309598/2020-6 and Funda\c{c}\~ao de Amparo \`a Pesquisa do Estado
do Rio Grande do Sul (FAPERGS), Grants No. 19/2551- 0000690-0 and No. 
19/2551-0001948-3.  D.S.R.~is supported by Funda\c{c}\~{a}o de
Amparo \`{a} pesquisa do estado de S\~{a}o Paulo - FAPESP
(Grant No. 2020/00560-0).  R.O.R. is partially supported by research grants from
CNPq, Grant No. 302545/2017-4, and Funda\c{c}\~ao Carlos Chagas Filho
de Amparo \`a Pesquisa do Estado do Rio de Janeiro (FAPERJ), Grant
No. E-26/201.150/2021.




\begin{thebibliography}{99}


\bibitem{Trodden:1998ym} M. Trodden, Electroweak baryogenesis,
  Rev. Mod. Phys. \textbf{71}, 1463 (1999).

\bibitem{Schwarz:2003du}
D. J. Schwarz, The first second of the universe,
Ann. Phys. (Berlin) \textbf{12}, 220 (2003).

\bibitem{BraunMunzinger:2009zz}
P. Braun-Munzinger and J. Wambach,
The phase diagram of strongly-interacting matter,
Rev. Mod. Phys. \textbf{81}, 1031 (2009).

\bibitem{Goldenfeld:1992qy}
N. Goldenfeld, Lectures on phase transitions and the renormalization group (CRC Press, Boca Raton,  1992).

\bibitem{Cornell:2002zz}
E. A. Cornell and C. E. Wieman,
Nobel Lecture: Bose-Einstein condensation in a dilute gas, the first 70 years and some recent experiments,
Rev. Mod. Phys. \textbf{74}, 875 (2002).

\bibitem{reentrant} K. C. Kao, \textit{Dielectric Phenomena in Solids} (Elsevier, New York, 2004).

\bibitem{Weinberg:1974hy}
S. Weinberg, Gauge and Global Symmetries at High Temperature,
Phys. Rev. D \textbf{9}, 3357 (1974).

\bibitem{Mohapatra:1979qt}
R. N. Mohapatra and G.~Senjanovic, Soft CP Violation at High Temperature,
Phys. Rev. Lett. \textbf{42}, 1651 (1979).

\bibitem{Klimenko:1988mb}
K. G. Klimenko, Gaussian effective potential and symmetry restoration at high temperatures in four-dimensional O($N$) X O($N$) field theory,
Z. Phys. C \textbf{43}, 581 (1989).

\bibitem{Bimonte:1995xs}
G. Bimonte and G. Lozano, Can symmetry nonrestoration solve the monopole problem?,
Nucl. Phys. \textbf{B460}, 155 (1996).

\bibitem{AmelinoCamelia:1996hw}
G. Amelino-Camelia, On the CJT formalism in multifield theories,
Nucl. Phys. \textbf{B476}, 255 (1996).

\bibitem{Orloff:1996yn}
J. Orloff, The UV price for symmetry nonrestoration,
Phys. Lett. B \textbf{403}, 309 (1997).

\bibitem{Roos:1995vm}
T. G. Roos, Wilson renormalization group study of inverse symmetry breaking,
Phys. Rev. D \textbf{54}, 2944 (1996).

\bibitem{Jansen:1998rj}
K. Jansen and M. Laine, Inverse symmetry breaking with 4-D lattice simulations,
Phys. Lett. B \textbf{435}, 166 (1998).

\bibitem{Bimonte:1999tw}
G. Bimonte, D. Iniguez, A. Tarancon, and C. L. Ullod, Inverse symmetry breaking on the lattice: An Accurate MC study,
Nucl. Phys. B \textbf{559}, 103 (1999)..

\bibitem{Pinto:1999pg}
M. B. Pinto and R. O. Ramos, A Nonperturbative study of inverse symmetry breaking at high temperatures,
Phys. Rev. D \textbf{61}, 125016 (2000).

\bibitem{Meade:2018saz}
P. Meade and H. Ramani, Unrestored Electroweak Symmetry,
Phys. Rev. Lett. \textbf{122},  041802 (2019).

\bibitem{Baldes:2018nel}
I. Baldes and G. Servant, High scale electroweak phase transition: Baryogenesis \& symmetry non-restoration, J. High Energy Phys. \textbf{10} (2018) 053.

\bibitem{Matsedonskyi:2020mlz}
O. Matsedonskyi and G. Servant, High-temperature electroweak symmetry non-restoration from new fermions and implications for baryogenesis,
J. High Energy Phys. \textbf{09}, (2020) 012. 

\bibitem{Matsedonskyi:2020kuy}
O. Matsedonskyi, High-temperature electroweak symmetry breaking by SM twins,
J. High Energy Phys. \textbf{04}, 036 (2021).

\bibitem{Bajc:2020yvd}
B. Bajc, A. Lugo, and F. Sannino, The free and safe fate of symmetry non-restoration,
Phys. Rev. D {\bf 103}, 096014 (2021).

\bibitem{Chai:2020zgq}
N. Chai, S. Chaudhuri, C. Choi, Z. Komargodski, E. Rabinovici, and M. Smolkin, Thermal order in conformal theories,
Phys. Rev. D \textbf{102},  065014 (2020).

\bibitem{Chaudhuri:2020xxb}
S. Chaudhuri, C. Choi, and E. Rabinovici, Thermal order in large N conformal gauge theories,
J. High Energy Phys. \textbf{04}, (2021) 203.

\bibitem{Chaudhuri:2021dsq}
S. Chaudhuri and E. Rabinovici, Symmetry breaking at high temperatures in large N gauge theories,
J. High Energy Phys.  \textbf{08},  (2021) 148.

\bibitem{Niemi:2021qvp}
L. Niemi, P. Schicho, and T. V. I. Tenkanen, Singlet-assisted electroweak phase transition at two loops, Phys. Rev. D \textbf{103}, 115035 (2021).

\bibitem{Ramazanov:2021eya}
S. Ramazanov, E. Babichev, D. Gorbunov, and A. Vikman, Beyond freeze-in: Dark matter via inverse phase transition and gravitational wave signal,
arXiv:2104.13722.


\bibitem{Maniv:2001zz}
T. Maniv, V. Zhuravlev, I. Vagner, and P. Wyder, Vortex states and quantum magnetic oscillations in conventional type-II superconductors,
Rev. Mod. Phys. \textbf{73}, 867 (2001).

\bibitem{Lai:2014nma}
D. Lai, Physics in very strong magnetic fields: Introduction and overview,
Space Sci. Rev. \textbf{191}, 13 (2015).

\bibitem{Andersen:2014xxa}
J. O. Andersen, W. R. Naylor, and A. Tranberg,
Phase diagram of QCD in a magnetic field: A review,
Rev. Mod. Phys. \textbf{88}, 025001 (2016).

\bibitem{Grasso:2000wj}
D. Grasso and H. R. Rubinstein, Magnetic fields in the early universe,
Phys. Rep. \textbf{348}, 163 (2001).

\bibitem{Shovkovy:2012zn}
I. A. Shovkovy, Magnetic catalysis: A review,
Lect. Notes Phys. \textbf{871}, 13 (2013).

\bibitem{Bandyopadhyay:2020zte}
A. Bandyopadhyay and R. L. S. Farias, Inverse magnetic catalysis: how much do we know about?,
Eur. Phys. J. ST \textbf{230}, 719 (2021).

\bibitem{Stevenson:1981vj}
P. M. Stevenson, Optimized perturbation theory,
Phys. Rev. D \textbf{23}, 2916 (1981).

\bibitem{Okopinska:1987hp}
A. Okopinska, Nonstandard expansion techniques for the effective potential in lambda phi**4 quantum field theory,
Phys. Rev. D \textbf{35}, 1835 (1987).

\bibitem{Klimenko:1992av}
K. G. Klimenko, Nonlinear optimized expansion and the Gross-Neveu model,
Z. Phys. C \textbf{60}, 677 (1993).

\bibitem{Kleinert:1998zz}
H. Kleinert, Strong coupling phi**4 theory in four epsilon dimensions, and critical exponents,
Phys. Rev. D \textbf{57}, 2264 (1998).

\bibitem{Chiku:1998kd}
S. Chiku and T. Hatsuda, Optimized perturbation theory at finite temperature,
Phys. Rev. D \textbf{58}, 076001 (1998).

\bibitem{Pinto:1999py}
M. B. Pinto and R. O. Ramos, High temperature resummation in the linear delta expansion,
Phys. Rev. D \textbf{60}, 105005 (1999).

\bibitem{Farias:2008fs}
R. L. S. Farias, G. Krein, and R. O. Ramos, Applicability of the linear delta expansion for the lambda phi**4 field theory at finite temperature in the symmetric and broken phases,
Phys. Rev. D \textbf{78}, 065046 (2008).

\bibitem{Yukalov:2019nhu}
V. I. Yukalov, Interplay between approximation theory and renormalization group,
Phys. Part. Nucl. \textbf{50}, 141 (2019).

\bibitem{Rosa:2016czs}
D. S. Rosa, R. L. S. Farias, and R. O. Ramos, Reliability of the optimized perturbation theory in the 0-dimensional $O(N)$ scalar field model,
Physica A \textbf{464}, 11 (2016).

\bibitem{Kneur:2007vj}
J. L. Kneur, M. B. Pinto, R. O. Ramos, and E. Staudt,
Updating the phase diagram of the Gross-Neveu model in $2+1$ dimensions,
Phys. Lett. B \textbf{657}, 136 (2007).

\bibitem{Kneur:2010yv}
J. L. Kneur, M. B. Pinto, and R. O. Ramos, Thermodynamics and phase structure of the two-flavor Nambu-Jona-Lasinio model beyond large-$N_c$,
Phys. Rev. C \textbf{81}, 065205 (2010).

\bibitem{Duarte:2011ph}
D. C. Duarte, R. L. S. Farias, and R. O. Ramos, Optimized perturbation theory for charged scalar fields at finite temperature and in an external magnetic field,
Phys. Rev. D \textbf{84}, 083525 (2011).

\bibitem{Duarte:2017zdz}
D. C. Duarte, R. L. S. Farias, P. H. A. Manso, and R. O. Ramos, Optimized perturbation theory applied to the study of the thermodynamics and BEC-BCS crossover in the three-color Nambu\textendash{}Jona-Lasinio model,
Phys. Rev. D \textbf{96}, 056009 (2017).

\bibitem{Kneur:2002kq}
J. L. Kneur, M. B. Pinto, and R. O. Ramos, Convergent Resummed Linear Delta Expansion in the Critical O(N) (phi**2(i))**2(3-d) model,
Phys. Rev. Lett. \textbf{89}, 210403 (2002).

\bibitem{Elizalde:1994gf}
E. Elizalde, S. D. Odintsov, A. Romeo, A. A. Bytsenko, and S. Zerbini, \textit{Zeta regularization techniques with applications} (World Scientific, Singapore, 1994).

\bibitem{Kapusta:2006pm}
 J. I. Kapusta and C. Gale, \textit{Finite-Temperature Field Theory:
Principles and Applications} (Cambridge University Press,
Cambridge, England, 2006).

\bibitem{Ayala:2004dx}
A. Ayala, A. Sanchez, G. Piccinelli, and S. Sahu, Effective potential at finite temperature in a constant magnetic field. I. Ring diagrams in a scalar theory,
Phys. Rev. D \textbf{71}, 023004 (2005).

\bibitem{Ayala:2014iba}
A. Ayala, M. Loewe, A. J. Mizher, and R. Zamora, Inverse magnetic catalysis for the chiral transition induced by thermo-magnetic effects on the coupling constant,
Phys. Rev. D \textbf{90}, 036001 (2014).


\end{thebibliography}
\end{document}